\documentclass[aps,prd,showkeys]{revtex4}
\usepackage{amsfonts}
\usepackage{amsmath}
\usepackage{amssymb}
\usepackage{bm}
\usepackage{graphicx}	
\usepackage{bm}

\def\sm(#1){{\scriptscriptstyle(#1)}}

\newcommand{\haf}{{\textstyle\frac{1}{2}}}
\newcommand{\PM}{{\hphantom{-}}}

\newcommand{\enn}{{p}}

\begin{document}

\title[Electromagnetic Wave in Gravitational-Wave Spacetime]{An Electromagnetic Plane Wave in the Spacetime of a Plane Gravitational Wave}

\author{C. R. Gwinn}
\email{cgwinn@ucsb.edu }
\affiliation{Physics Department, University of California, Santa Barbara 93106, USA}

\date{\today}

\begin{abstract}
I find nearly plane-wave solutions for the Gauss-Ampere law for the 4-vector potential, subject to the Lorenz gauge condition, in the spacetime of a plane gravitational plane wave.
I assume that the gravitational wave is weak, in the sense that the dimensionless strain amplitude $h$ is much less than 1.
I find a solution for the homogeneous scalar wave equation in this spacetime, and then find a 4-vector potential that solves the Gauss-Ampere law and Lorenz gauge condition in the absence of sources.
The solutions are plane waves in Minkowski spacetime, plus additional scattered waves of order $h$.
The problem is analogous to diffraction from a transmission grating, or Brillouin scattering from sound waves in matter.
The corrections solve the inhomogeneous wave equation in Minkowski spacetime, with a ``distributed source'' of order $h$ comprised of terms arising from the non-Minkowski metric and the zero-order solution.
The scalar wave solution requires two scattered waves, which can be combined to form a phase correction $h\tilde \varphi$ that varies at the gravitational-wave frequency.
This phase correction yields the same time delay and deflection at the observer as for propagation along null geodesics, in the ray approximation.
The electromagnetic-wave solution requires four scattered waves. Two correspond to the phase correction $h\tilde \varphi$ found for the scalar field.
The other two scattered waves introduce amplitude and polarization changes of order $h$ to the electromagnetic wave.
The time delay and deflection match those for the scalar waves. The solution predicts variations of the intensity of the electromagnetic wave of first order in $h$, at the wavenumber of the gravitational wave.
These arise from interference of the first-order scattered waves and the zero-order solution. I briefly discuss possible observations of this effect.
\end{abstract}


\maketitle

\section{Introduction}\label{sec:Introduction}

Propagation of electromagnetic waves along geodesics in the spacetime of a gravitational wave has been extensively studied \cite{EstabrookWahlquist1975GReGr...6..439E,Burke1975ApJ...196..329B,Detweiler1979ApJ...234.1100D,HellingsDowns1983ApJ...265L..39H,Pyne1996,Kopeikin99,BookFlanagan11}. 
These studies have focused on finding the null geodesics in this spacetime, and then determining the observed delay and deflection for propagation along these paths.
Linder \cite{Linder1988ApJ...326..517L,Linder1988ApJ...328...77L} found amplitude variations of order $h^2$ from the path geometry.
Kopeikin and collaborators extended these calculations to find the polarization and amplitude of light, as well as delay and deflection, after propagation through wave and static gravitational fields from arbitrary sources
\cite{1999PhRvD..60l4002K,Kopeikin99,2002PhRvD..65f4025K,2006CQGra..23.4299K,2014frc1.book.....K}.
However, electromagnetic radiation is fundamentally a wave. The geodesic is an approximation to the wave equation, although often an excellent one. 
Here I consider propagation of electromagnetic plane waves, in the spacetime of a gravitational plane wave, from the standpoint of Maxwell's equations, 
or more precisely from the Gauss-Ampere law and the Lorenz gauge for the 4-vector potential.

Gravitational waves that might observably deflect or delay light signals arise as radiation by objects of stellar mass or greater with high acceleration, or as a gravitational-wave background from the early universe
(see, for example, \cite{MTW}, Chapter 36).
The dimensionless strain $h$ gives the amplitude of the gravitational wave, and the frequency $K$, or equivalently the wavenumber, gives its time and space variation. 
The spectrum of gravitational waves in the vicinity of Earth is poorly known.
However, extensive theoretical work, upper limits, and recent detections suggest that for frequencies $K/2\pi$ over the range of a few Hz down to $10^{-9}\ Hz$ and less, the strain may commonly be $h\sim 10^{-21}$, although it may reach as much as $10^{-17}$.
For longer periods, the strain may be greater, but much less than 1.
Frequencies of electromagnetic waves observable from Earth and its environs range upward from $k/2\pi \approx 10^6\ \mathrm{Hz}$, although observations from Earth's surface are limited by the ionosphere to $k/2\pi \gtrsim 10^7\ \mathrm{Hz}$.
Consequently, in this work, I will assume that $h\ll 1$, and discard terms of order $h^2$ or higher power.
I also note that $K \ll k$, although I will not use that fact to discard terms.

\subsection{Phase, delay, and apparent direction}\label{sec:PhaseDelayDirectionIntro}

Effects of propagation of electromagnetic-waves through gravitational waves are a delay in the arrival time of a signal, and a change in the apparent direction.
Changes of the observed delay, and the apparent direction of the source, are related to the change in phase caused by the gravitational wave.
Delay is the derivative of phase with respect to frequency (see \cite{Jackson}, Chapter 7):
\begin{align}
\label{eq:TauEqDPhiDk}
\tau &= \partial_k \Phi
\end{align}
where $\tau$ is the delay, $k$ the angular frequency, and  $\Phi$ the phase.
The fact that the Fourier transform of a delayed wave packet is the Fourier transform of the undelayed packet, multiplied by the product $\tau k$, demonstrates Equation\ \ref{eq:TauEqDPhiDk}.

The apparent direction of the source $\bm{\hat s}$ is a spatial unit 3-vector, opposite to the normal to the wave.
The spatial 3-gradient of phase is opposite $\bm{\hat s}$. The 4-gradient of phase is a null vector for a plane wave, or one that is locally approximately a plane wave, so the time component serves to normalize $\bm{\hat s}$.
\begin{align}
\label{eq:sEqGradPhi}
\hat s_j&= -\frac{1}{\left|\partial_t \Phi\right|} \partial_j \Phi
\end{align}

If several plane waves are present, they can nevertheless locally sum to a single plane wave, if the phase relationship among components with different polarization, frequency, and wavenumber remains nearly constant over the size of the observer's sampling region.
For electromagnetic waves, the Poynting 3-vector provides a simple means of forming such a sum.
The apparent source direction is the unit vector opposite the Poynting vector.
Waves that are not locally plane waves may involve multiple delays and multiple directions, and require further dissection.

Measurement of delay requires that the signal have some degree of temporal coherence, and measurement of direction requires that the signal have at least some spatial coherence,
in the region of spacetime near the observer.

Gravitational waves can also affect the amplitude and polarization of electromagnetic waves, as calculated in this work.

\subsection{Outline of the paper}

I seek a form for plane-like electromagnetic waves in the spacetime of a weak plane gravitational wave.
This gravitational wave perturbs Minkowski spacetime. The fields of the electromagnetic wave must solve the Maxwell equations, and must satisfy the Lorenz gauge condition.
In Section\ \ref{sec:GravWave}, I introduce the metric and connection for such a gravitational wave of the ``+'' polarization, with dimensionless strain amplitude $h \ll 1f$, and with frequency and wavenumber $K$ traveling toward $+z$.
The wave produces corrections to the metric $h \tilde w$, where $\tilde w=\cos[K(-t+z)]$.
Gravitational plane waves with other polarizations and directions can be constructed from superpositions and rotations of this wave.
I find the connection and note that it takes a highly symmetric form, and show that the curvature is zero, through first order in $h$.

In Section\ \ref{sec:WaveEqScalarField}, I find the homogeneous wave equation for a scalar field. For a scalar plane wave $\Psi_0[x^\alpha]$, the gravitational wave effectively introduces a ``distributed source'' of order $h \Psi_0$, 
which demands two correcting waves of order $h$.
Because the correcting waves are of order $h$, it need solve only the inhomogeneous wave equation for Minkowski spacetime, with that ``distributed source''.
I assume a plane-wave form for  $\Psi_0$, and find a nearly plane-wave solution to that inhomogeneous wave equation.
For a scalar plane wave, the correction can take the equivalent forms of additional waves, or of a change of phase.
I point out the analogy of the process to diffraction by a transmission grating, or Brillouin scattering from sound waves in a solid. 
The calculated phase change $\tilde \varphi$ is the same as that inferred for propagation along a geodesic through the same spacetime.
This phase change is of order $h k/K$.

I extend the discussion to electromagnetic waves in Section\ \ref{sec:EMinGWspacetime}.
Electromagnetic waves are fully characterized by the 4-vector potential $a_\alpha$.
This potential must satisfy the Gauss-Ampere law, without charges or currents,
and the Lorenz gauge condition.
I expand these conditions through first order in $h$, in the absence of charges or currents.
To zeroth order in $h$, the spacetime is Minkowski spacetime,
and the Gauss-Ampere law and Lorentz gauge condition imply a homogeneous wave equation for each component of the zero-order solution, $a^{\sm(0)}_{\ \ \alpha}$, as is well known. 
The expressions through first order in $h$ require a correcting potential $a^{\sm(1)}_{\ \ \alpha}$.
This correction satisfies the wave equation for Minkowski spacetime in each component, with a ``distributed source'' involving the zero-order solution $a^{\sm(0)}_{\ \ \alpha}$ times $h$.
Corrections proportional to $h\tilde w$ have the same form as those for the scalar wave equation for each component of $a^{\sm(1)}_{\ \ \alpha}$, and thus yield the same additional scattered waves, or phase change $\tilde \varphi$, as for the scalar wave equation.
Corrections proportional to derivatives of $\tilde w$ combine the components of $a^{\sm(1)}_{\ \ \alpha}$, and so lead to correcting waves that create 
polarization and amplitude variations. The effects of these corrections is smaller than those proprtional to $\tilde w$ by a factor of $K/k$.

In Section\ \ref{sec:NearlyPlaneEMWaves},
I define the two independent polarizations for plane electromagnetic waves in Minkowski spacetime,
transverse electric (TE) and transverse magnetic (TM).
These serve as the zero-order solutions of the Gauss-Ampere law and Lorenz gauge condition $a^{\sm(0)}_{\ \ \alpha}$. 
I find the field tensor for these waves.
Motivated by the zero-order solution, the inhomogeneous wave equation and Lorenz gauge condition, the solution for the scalar wave equation,
and the forms of the corrections to the zero-order solution that form the effective ``distributed source'',
I propose a candidate solution for $a^{\sm(0)}_{\ \ \alpha}$ involving four waves, scattered from the zero-order electromagnetic wave by the gravitational wave.
I show that, with appropriate parameters, this candidate solution satisfies the Gauss-Ampere law and the Lorenz gauge condition, through first order in $h$.
The two scattered waves corresponding to a phase change is the same as that found for the scalar field in Section \ \ref{sec:WaveEqScalarField}, $\tilde \varphi$.
The two other scattered waves correspond to amplitude and polarization changes.
Again, these wave result from diffraction and are not described by goemetric optics.
I note that these solutions are not unique:
any solution that satisfies the Gauss-Ampere law and the Lorenz gauge condition can be added to produce another solution.

In Section\ \ref{sec:DelayDirectionScalar}
I calculate the delay and the apparent deflection of the wave at an observer, from the phase correction.  These observables match the results of previous geodesic-based calculations.
In Section\ \ref{sec:DelayDirectionEM},
I calculate the field tensor for the TE polarization, and observe that the field tensor for the TM polarization is related by a duality transformation.
(This is the case for the potentials explored here, but need not be the case for independent polarizations, in general).
I find that the delay, as calculated from the phase, and the wave direction, as calculated from the normalized Poynting vector,
agreee with the results for the scalar wave.
The amplitude of the wave is changed by the amplitude and polarization changes introduced by scattering.
This change produces local variations of the wave intensity, but does not affect the intensity averaged over many periods or wavelengths of the gravitational wave.
This amplitude variation results from interference of the scattered waves with the zero-order solution.

I briefly discuss possible observational tests of the predicted variations of electromagnetic-wave intensity and the other observables in Section\ \ref{sec:Discussion},
and summarize the results in Section\ \ref{sec:Discussion}.

\section{Spacetime of a gravitational wave}\label{sec:GravWave}

\subsection{Coordinates and metric for a gravitational-wave spacetime}\label{sec:GravWaveDefs}

I follow the description of Pyne et al. \cite{Pyne1996} for gravitational-wave spacetime, but omit the cosmological effects included in that paper.
I set $c=1$, and use Greek letters for 4-indices $\{ t, x, y, z \}$ and Latin for 3-indices $\{x, y, z\}$.
I suppose that the gravitational wave is a plane wave with the ``$+$'' polarization,
and dimensionless strain amplitude $h$. I suppose that $h\ll 1$.
I use Cartesian coordinates:
\begin{align}
x^\beta &= \left( t, x, y, z \right) 
\end{align}
The gravitational wave is polarized along $x$ and $y$ axes (the ``$+$'' polarization), and travels toward $\hat z$.
The metric is thus (\cite{MTW}, Section 35.6):
\begin{align}
\label{eq:GWMetric}
g_{\alpha\beta} &=
\begin{pmatrix}
-1 & 0 & 0& 0 \\
0 & 1+ h \tilde w & 0 & 0 \\
0 & 0 & 1 - h \tilde w & 0 \\
0 & 0 & 0 & 1
\end{pmatrix}
.
\end{align}
Here,
the function:
\begin{align}
\tilde w &= \cos\left( K_\beta x^\beta \right)
\end{align}
gives the time and space variation of the gravitational wave.
The 4-wavevector of the gravitational wave is:
\begin{align}
K_\beta &=( -K, 0, 0, K)
\end{align}
where $K$ is the wavenumber of the gravitational wave.
Without loss of generality, I have taken the spatial direction of the gravitational wave as $\hat z$.
This metric is identical to that of Pyne et al. \cite{Pyne1996}, with their rotation matrix $R$ set equal to the identity matrix.

It is convenient to write:
\begin{align}
g_{\alpha\beta} &= \eta_{\alpha\beta} + h \tilde w e^{\sm(+)}_{\, \alpha\beta} 
\end{align}
where $\eta_{\alpha\beta}$ is the usual metric for Cartesian coordinates in Minkowski spacetime, equal to $g_{\alpha\beta}$ in the limit $h\rightarrow 0$.
The polarization matrix for the gravitational wave is:
\begin{align}
e^{\sm(+)}_{\, \alpha\beta}  &=
\begin{cases}
+1 & \alpha=\beta=x \\
-1 & \alpha=\beta=y \\
0 & \mathrm{otherwise}
\end{cases}
.
\end{align}
The symbol $e^{\sm(+)}_{\, \alpha\beta}$ is not a covariant tensor.
In particular, the analogous exprssion for contravariant indices is $e^{\sm(+)\, \alpha\beta} = - e^{\sm(+)}_{\, \alpha\beta}$,
which appears in the contravariant form of the metric:
\begin{align}
g^{\alpha\beta} &= \eta^{\alpha\beta} + h \tilde w e^{{\sm(+)} \, \alpha\beta} 
.
\end{align}

\subsection{Connection}

Because the metric is not Minkowski, covariant derivatives appear in the wave equation \cite{MTW}.
In general, these derivatives involve the metric and the connection, given by the Christoffel symbols.
For the metric of a gravitational wave given by Equation\ \ref{eq:GWMetric},
through first order in $h$, the nonzero Christoffel symbols of the second kind are:
\begin{alignat}{5}
\label{eq:Christoffel}
&\haf h K \sin\left[ K_\lambda x^\lambda \right]
  &  =- \Gamma^t_{xx} &&=\PM   \Gamma^t_{yy} &&= -     \Gamma^z_{xx} &&= \PM \Gamma^z_{yy}  \\
&& =- \Gamma^x_{tx} &&=\PM  \Gamma^x_{zx} &&=\PM \Gamma^y_{ty} &&=       - \Gamma^y_{zy} \nonumber \\
&& =- \Gamma^x_{xt} &&=\PM  \Gamma^x_{xz} &&=\PM \Gamma^y_{yt} &&=       - \Gamma^y_{yz}  \nonumber
.
\end{alignat}
The last line follows from the second line and the symmetry of the Christoffel symbols $\Gamma^\alpha_{\beta\gamma} = \Gamma^\alpha_{\gamma\beta}$.
All other Christoffel symbols of the second kind are zero.
Clearly, this particular set of Christoffel symbols has a great deal of symmetry. 
In particular, for any $\alpha$,
\begin{align}
\label{eq:ChristoffelFact1}
\Gamma^\kappa_{\kappa\alpha}&=0
\end{align}
where only $\kappa=x$ and $\kappa=y$ contribute to the sum.
For any $\alpha$, $\beta$, $\gamma$,
\begin{align}
\partial_x \Gamma^\gamma_{\alpha\beta}&=\partial_y \Gamma^\gamma_{\alpha\beta}=0 
.
\end{align}
Furthermore, because $\partial_t \sin\left[ K_\lambda x^\lambda \right] = -\partial_z \sin\left[ K_\lambda x^\lambda \right]$, and $\partial_x \sin\left[ K_\lambda x^\lambda \right]=\partial_y \sin\left[ K_\lambda x^\lambda \right]=0$,
\begin{align}
\partial_\kappa \Gamma^\kappa_{\alpha\beta}&=0 
\end{align}
for all $\alpha$, $\beta$. 

The curvature tensors express the intrinsic curvature of spacetime \cite{MTW}.
The Riemann curvature tensor is
\begin{align}
R^\rho_{\sigma\mu \nu} &= \partial_\mu \Gamma^\rho_{\nu\sigma} - \partial_\nu \Gamma^\rho_{\mu\sigma} + \Gamma^\rho_{\lambda\mu} \Gamma^\lambda_{\mu\tau} - \Gamma^\rho_{\lambda\nu} \Gamma^\lambda_{\sigma\mu} 
.
\end{align}
In the present case, the products of Christoffel symbols can be ignored because each Christoffel symbol is already first order in $h$.
The Ricci curvature tensor is then:
\begin{align}
R_{\sigma\mu} &= R^\kappa_{\sigma\mu \kappa} =  \partial_\mu \Gamma^\kappa_{\kappa\sigma} - \partial_\kappa \Gamma^\kappa_{\mu\sigma} = 0
,
\end{align}
to first order in $h$,
and similarly 
\begin{align}
\label{eq:Ricci11}
R^{\tau}_{\mu} &= g^{\tau\sigma} R_{\sigma\mu} =0 .
\end{align}

\begin{figure}
\centering
\includegraphics[width=0.42\textwidth]{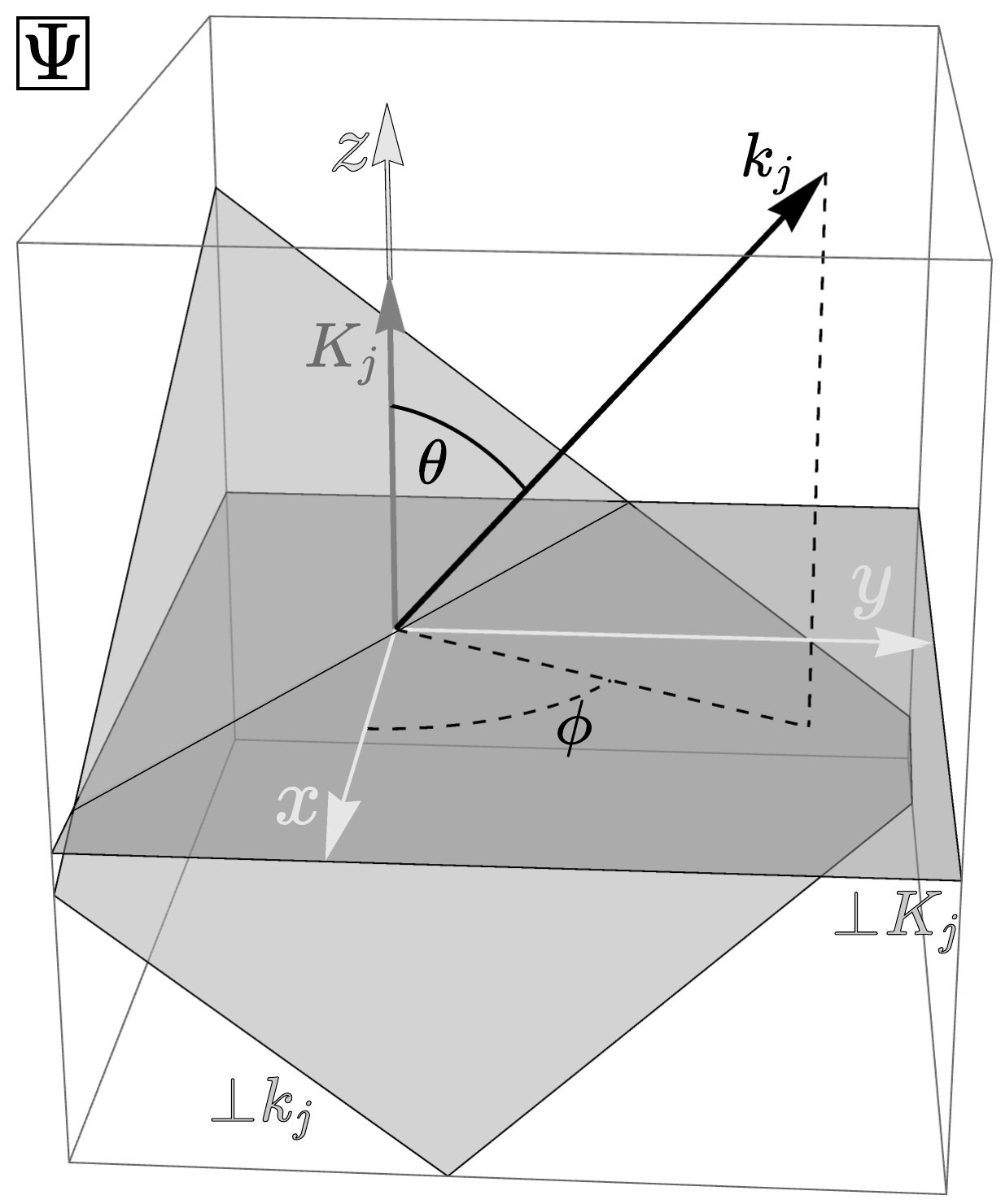} %
\caption{
\label{fig:WaveVectorDiagramScalar}
Spatial geometry for scalar waves in gravitational-wave spacetime,
showing the spatial wavevectors for the gravitational wave $K_j$ and the scalar-field wave $k_j$.
The planes perpendicular to $K_j$ and $k_j$ are surfaces of constant phase for the two waves.
The diagram is in the limit of $h\rightarrow 0$.
}
\end{figure}

\section{Waves of a scalar field in gravitational-wave spacetime}\label{sec:WaveEqScalarField}

Consider a scalar field $\Psi [x^\alpha ]$, in the gravitational-wave spacetime and coordinates set forth in Section\ \ref{sec:GravWaveDefs}.
The homogeneous scalar wave equation states:
\begin{align}
\Psi^{;\gamma}_{\ \ \gamma}  &=0
\end{align}
where a semicolon denotes a covariant derivative. In this equation, 
$R^\gamma_{\ \gamma}$ is the contraction of the Ricci curvature tensor, equal to zero to first order in $h$ for this spacetime.
Explicitly, the covariant derivatives on the left side of this equation are:
\begin{align}
\Psi^{;\gamma}_{\ \ \gamma}  &=\partial_\gamma \left(  g^{\gamma\alpha} \partial_\alpha \Psi \right) + \Gamma^\gamma_{\gamma \kappa}  \left( g^{\kappa\alpha} \partial_\alpha \Psi \right) 
.
\end{align}
As Equation\ \ref{eq:ChristoffelFact1} states, $\Gamma^\gamma_{\gamma \kappa} =0$.
The homogeneous scalar wave equation thus takes the simple form:
\begin{align}
\partial_\gamma \left(  g^{\gamma\alpha} \partial_\alpha \Psi \right) &=0
.
\end{align}
Moreover,
$g^{\gamma\alpha}$ depends only on $t$ and $z$, and is diagonal, and only $g^{xx}$ and $g^{yy}$ vary.
Consequently, 
\begin{align}
\label{eq:ScalarWaveEqInhomogForm}
\partial_\gamma  \eta^{\gamma\alpha} \partial_\alpha \Psi 
&= - h \tilde w e^{\sm(+)\, \alpha\beta} \partial_\alpha\partial_\beta \Psi \\
&=   h \tilde w  \left( \partial_{xx} - \partial_{yy} \right) \Psi
,
\nonumber
\end{align}
where the second line follows from the fact that $e^{\sm(+)\, \alpha\beta} \partial_\alpha \tilde w = 0$.

One approach to solving Equation\ \ref{eq:ScalarWaveEqInhomogForm} is to take note of the coefficient $h$ on the right, and suppose that $\Psi$ consists of a zero-order solution $\Psi^{\sm(0)}$ plus a first-order solution $h \Psi^{\sm(1)}$.
Equation\ \ref{eq:ScalarWaveEqInhomogForm} then becomes the two coupled equations:
\begin{align}
\label{eq:ScalarWaveEqInhomogFormPart0}
\partial_\gamma  \eta^{\gamma\alpha} \partial_\alpha \Psi^{\sm(0)} &= 0 \\ 
\label{eq:ScalarWaveEqInhomogFormPart1}
h \partial_\gamma  \eta^{\gamma\alpha} \partial_\alpha \Psi^{\sm(1)} &=   h \tilde w  \left( \partial_{xx} - \partial_{yy} \right) \Psi^{\sm(0)} 
.
\end{align}
Given a solution of Equation\ \ref{eq:ScalarWaveEqInhomogFormPart0} for $\Psi^{\sm(0)}$, the problem consists of finding a solution to the inhomogeneous wave equation\ \ref{eq:ScalarWaveEqInhomogFormPart1} for $h \Psi^{\sm(1)}$.
A modification of this approach is to suppose that the first-order solution appears as an exponential term
\begin{align}
\label{eq:WavesOrPhase}
\Psi &= \Psi^{\sm(0)} e^{i h \tilde \varphi} = \Psi^{\sm(0)} + i h \, \tilde \varphi \Psi^{\sm(0)} + {\mathcal O}\left[ h^2 \right].
\end{align}
One can regard the unknown function $\tilde \varphi$ as an additive phase, or as a change in amplitude if $\tilde \varphi$ is imaginary.
I will use these expressions interchangeably.

\begin{figure}
\centering
\includegraphics[width=0.42\textwidth]{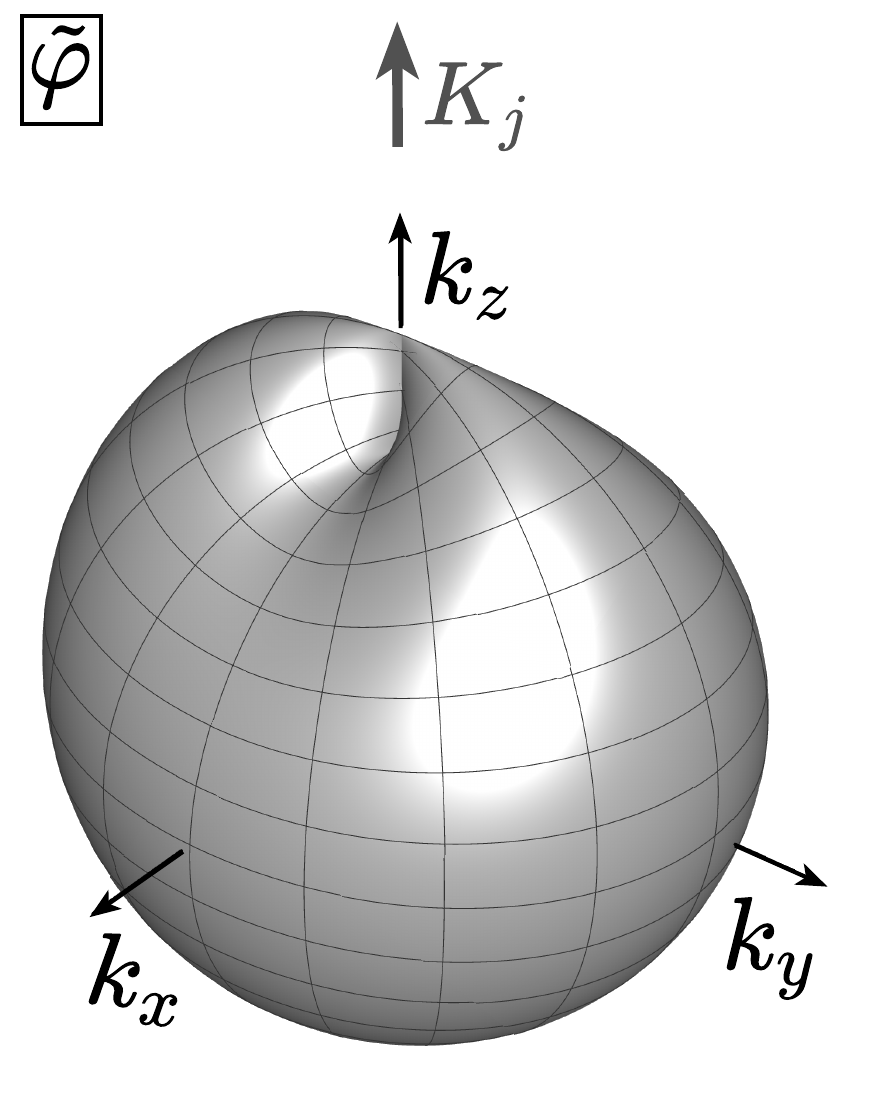} %
\caption{
\label{fig:PhasePattern}
The phase change due to the gravitational wave $\tilde\varphi$, shown as a function of the spatial components of the wavevector, $k_j$, at the spatial origin at time $t=0$.
The surface shows the change of phase as a departure from a sphere centered at $k_j=0$, with negative phases shown as a decrease of radius and positive as an increase.
}
\end{figure}

I assume that $\Psi^{\sm(0)}$ takes the form of a plane wave:
\begin{align}
\label{eq:ScalarFieldDefinition}
\Psi &= \Psi^{\sm(0)} e^{i h \tilde \varphi} 
= a e^{ i k_\alpha x^\alpha + i h \tilde \varphi }
\end{align}
with the null wavevector:
\begin{align}
k_\alpha &= \{ k_t, k_x, k_y, k_z \} = \{ -k, k \sin\theta\cos\phi, k \sin\theta\sin\phi, k\cos\theta \}
.
\end{align}
where $\{\theta,\phi\}$ are the usual spherical coordinates.
This $\Psi^{\sm(0)}$ solves the scalar wave equation in Minkowski spacetime,\ \ref{eq:ScalarWaveEqInhomogFormPart0}.

I expect that $\tilde \varphi$ depends only on $t$ and $z$, because the problem is independent of translation in $x$ and $y$, aside from an overall phase factor for $\Psi^{\sm(0)}$.
The inhomogeneous scalar wave equation\ \ref{eq:ScalarWaveEqInhomogFormPart1} becomes: 
\begin{align}
\label{eq:ScalarWaveEqPhase}
\partial_\gamma  \eta^{\gamma\alpha} \partial_\alpha \left( a \exp\left[ i k_\alpha x^\alpha \right] i h \tilde \varphi \right)
&= - h a k^2              \sin^2\theta \cos[2\phi]  \left( e^{ i k_\alpha x^\alpha }  \cos[K(-t+z) ]  \right)
.
\end{align}
The time and space variation on the right side is the sum of two complex waves:
\begin{align}
\label{eq:WaveSum}
e^{ i k_\alpha x^\alpha } \cos[K(-t+z) ]  &= \haf \left( e^{ i (k_\alpha+K_\alpha) x^\alpha } + e^{ i (k_\alpha-K_\alpha) x^\alpha } \right)
\end{align}
These can be regarded as additional, scattered waves produced by interaction of the gravitational and scalar waves.
Their 4-wavevectors are the sum and difference of those of the scalar and electromagnetic waves.
Their 4-wavevectors $(k_\alpha+K_\alpha)$ and $(k_\alpha-K_\alpha)$ are not null. 
(Unless, of course, the spatial components of $k_\alpha$ and $K_\alpha$ are parallel.)
If the wavevectors of these two waves were null, then application of the Minkowski-space d'Alembertian would yield 0,
and they could not contribute to solution of Equation\ \ref{eq:ScalarWaveEqPhase}.
Nevertheless, they are eigenfunctions of the d'Alembertian.
Application of the d'Alembertian to the two exponentials on the right side of Equation\ \ref{eq:WaveSum} leaves their forms unchanged, while multiplying by factors $\mp 2 k K (1-\cos\theta)$ for $\pm K_\alpha$.

To account for the different signs the eigenvalues of the two exponentials,
I change cos to sine:
\begin{align}
e^{ i k_\alpha x^\alpha }  \sin[K(-t+z) ] &= \frac{1}{2 i} \left( e^{ i (k_\alpha+K_\alpha) x^\alpha } - e^{ i (k_\alpha-K_\alpha) x^\alpha } \right) \nonumber
\end{align}
 I find that this combination yields:
 \begin{align}
 \label{eq:DAlembertianOfExpSineCombo}
\partial_\gamma  \eta^{\gamma\alpha} \partial_\alpha \left(   \exp[ i k_\alpha ] \sin[K(-t+z)] \right) 
&=- 2 i k K (1-\cos\theta)  \exp[ i k_\alpha ] \cos[K(-t+z)]
.
\end{align}
Comparison of the right side of Equation\ \ref{eq:ScalarWaveEqPhase} with that of Equation\ \ref{eq:DAlembertianOfExpSineCombo}
yields $\tilde \varphi$:
\begin{align}
\label{eq:ScalarVarphidef}
i h \, \tilde\varphi[t,z] %
&=i h \left(  -  \frac{k}{2 K}(1+\cos\theta) \cos(2\phi ) \sin[ K (-t + z)] \right)
.
\end{align}
With this choice of $\tilde\varphi$, $\Psi$ as given by Equation\ \ref{eq:ScalarFieldDefinition} solves the wave equation\ \ref{eq:ScalarWaveEqInhomogForm}.
Figure\ \ref{fig:PhasePattern} displays the function $\tilde\varphi$ at the origin as a function of the direction of the spatial wavevector $k_j$.

As discussed in Section\ \ref{sec:PhaseDelayDirectionIntro}, the delay observed for a wave packet is:
\begin{align}
\tau &= \partial_k h \tilde \varphi 
.
\end{align}
Thus, the phase $\tilde \varphi$ in Equation\ \ref{eq:ScalarVarphidef} is in agreement with calculation of delay $\tau$ along the perturbed geodesic in this spacetime ([see, for example \cite{Pyne1996}):
\begin{align}
\tau &=  - h \frac{1}{2 K}(1+\cos\theta) \cos(2\phi ) \sin[ K (-t + z)] 
.
\end{align}

In terms of the discussion of observable delay and direction in Section\ \ref{sec:PhaseDelayDirectionIntro},
interaction with the gravitational wave has produced two additional, scattered waves.
These waves do not have null wavevectors, because they are modified by the ``extended source'' of Equation\ \ref{eq:ScalarWaveEqInhomogFormPart1} at each event throughout spacetime.
Together, the three waves solve the scalar wave equation.
The observer must deal with the sum of these waves, to determine delay and direction.
The phase differences among the three waves vary on space or time scales comparable to $1/K$,
so superposition of the waves is possible so long as the observer's instrument and observing time are smaller than that.

The creation of additional, scattered waves is analogous to the action of a diffraction grating used in transmission, 
or still more closely to Brillouin scattering of light from sound waves in matter \cite{Wolff:21}.
Just as with interaction of a scalar wave with a gravitational wave,
the interaction of a light wave with a periodic structure or sound waves produces scattered waves, at the sum and difference wave vectors.
For Brillouin scattering, the sum and difference waves do not solve the wave equation within the material, where they are supported by a distributed source, as for the gravitational-wave scattering discussed here.
For both the transmission grating and Brillouin scattering, unlike the gravitational-wave case, observations are made outside the scatterer, and conditions at the boundary define a transition to propagation in free space.
In all cases, the diffracting structures are far larger than the wavelength of the scattered wave.
In the case of a scalar field scattered by a gravitational wave, the construction of Equation\ \ref{eq:WavesOrPhase} allows one to easily superpose the three waves,
and treat the effect of the gravitational wave as a simple phase change.

\section{Electromagnetism in gravitational-wave spacetime}\label{sec:EMinGWspacetime}

In this section, I find expressions for the laws of electromagnetism in the spacetime introduced in the previous section.
I express the electromagnetic fields in terms of the 4-vector potential $a_\alpha$ (see \cite{MTW} Section 22.4).
Gauss's law for the magnetic field and Faraday's law are satisfied automatically by the use of this potential.
The problem then becomes that of solving Gauss's law and Ampere's law including displacement current,
subject to the Lorenz gauge condition.
Here, Gauss's law and Ampere's law are expressed as the Gauss-Ampere law for the electromagnetic field tensor.

\subsection{Lorenz gauge condition and Gauss-Ampere law}

Electromagnetic waves are solutions of the Gauss-Ampere law.
Use of the potentials automatically satisfy Gauss's law for magnetism and Faraday's law.
The Gauss-Ampere law takes a convenient form when we demand that $a_\alpha$ satisfy the Lorenz gauge condition:
\begin{align}
\label{eq:LorenzGaugeConditionContravariant}
a^\alpha_{\ ;\alpha} &= 0
\end{align}
I take this approach here.

In curved spacetime, just as in Minkowski spacetime,
the covariant vector potential $a_\alpha$ is related to the covariant field tensor $F_{\mu\nu}$ via (see \cite{MTW}, Problem 22.8):
\begin{align}
\label{eq:covFab}
F_{\mu\nu} &= a_{\nu,\mu} -a_{\mu,\nu}
\end{align}
where the Christoffel symbols cancel, so that the covariant derivatives are simply partial derivatives in this case.
When the Lorenz gauge condition is satisfied, 
the Gauss-Ampere law takes the general form:
\begin{align}
\label{eq:GaussAmpere}
F^{\alpha\beta}_{\ \ \ ;\beta} &= 4 \pi J^\alpha - R^{\alpha}_{\ \,\beta} a^\beta
.
\end{align}
where 
$J^\alpha$ is the 4-current density and $R^{\alpha}_{\ \, \beta}$ is the Ricci curvature tensor.
As Equation\ \ref{eq:Ricci11} states, $R^{\alpha}_{\ \,\beta}=0$ for gravitational-wave spacetime.
I seek solutions far from any source charges or currents, so I assume:
\begin{align}
J^\alpha =0
,
\end{align}
and the Gauss-Ampere law takes the homogeneous form:
\begin{align}
F^{\alpha\beta}_{\ \ \ ;\beta} &= 0 \nonumber
\end{align}
or, with the explicit connection coefficients:
\begin{align}
\label{eq:GaussAmpereGW}
 \partial_\beta F^{\alpha\beta} + \Gamma^\alpha_{\beta\kappa} F^{\kappa \beta} + \Gamma^\beta_{\beta\kappa} F^{\alpha \kappa}  &= 0
. 
\end{align}
The field tensor is antisymmetric: $F^{\alpha\beta} = - F^{\beta\alpha}$. However, for gravitational-wave spacetime with coordinates specified in Section\ \ref{sec:GravWaveDefs}, 
the Christoffel symbol is symmetric under interchange of the lower two indices: $\Gamma^\alpha_{\beta\kappa} = \Gamma^\alpha_{\kappa\beta}$.
Consequently, the second term on the left side is its own negative under interchange of the dummy indices $\beta$ and $\kappa$, and so is zero.
The third term on the right is also zero for this spacetime because $\Gamma^\beta_{\beta\kappa}=0$, as Equation\ \ref{eq:ChristoffelFact1} states.
The Gauss-Ampere law for the gravitational-wave spacetime considered in this work is thus:
\begin{align}
\label{eq:GaussAmpereSimplified}
\partial_\beta F^{\alpha\beta} &=0
.
\end{align}

One can combine Equations\ \ref{eq:LorenzGaugeConditionContravariant},\ \ref{eq:covFab}, and\ \ref{eq:GaussAmpere} to find an equation for the vector potential.
Interestingly, the general result takes the simple form (\cite{MTW} Equation 22.19d):
\begin{align}
\label{eq:WaveEquationForVectorPotentialGW}
a^{\alpha ; \beta}_{\ \ \ \beta} =4\pi J^\alpha - R^\alpha_{\ \beta}a^\beta
.
\end{align}
This form is closely analogous to the Minkoski-space form.
However, in the following sections, I will follow the first approach above, using Equations\ \ref{eq:LorenzGaugeConditionContravariant},\ \ref{eq:covFab}, and \ref{eq:GaussAmpereSimplified}. 

Perhaps it is unnecessary to state that the resulting solutions are not unique.
For example, one can add any solution to the Minkowski-space wave equation with amplitude of order $h$, with arbitrary polarization and direction, and obtain another solution that has the same form in the limit $h\rightarrow 0$.
Furthermore, any difference between the solution presented here, and a different solution, is also a solution.

\subsection{Expansion through first order in $h$}

\subsubsection{Lorenz gauge condition}\label{sec:LorenzGaugeConditionExpand}

The Lorenz gauge condition for the 4-vector potential $a^\alpha$ is:
\begin{align}
0 &= a^\alpha_{\ \, ;\alpha} = \partial_\alpha a^\alpha + \Gamma^\alpha_{\alpha\beta}  a^\beta 
\end{align}
For gravitational-wave spacetime with the specified coordinates, $ \Gamma^\alpha_{\alpha\beta} =0$ as stated in Equation\ \ref{eq:ChristoffelFact1}, so that the term involving the connection vanishes.
Thus,
\begin{align}
0 
&= g^{\alpha\beta} \partial_\alpha a_\beta \nonumber
,
\end{align}
where I note that $\partial_\alpha g^{\alpha\beta}=0$ for the metric.
Thus,
\begin{align}
\label{eq:LorenzGaugeConditionCovariant}
\eta^{\alpha\beta} \partial_\alpha a_\beta &= - h \tilde w e^{\sm(+)\, \alpha\beta}  \partial_\alpha a_\beta
.
\end{align}
As this equation suggests, I seek solutions that the sum of a Minkowski-space solution $a^{\sm(0)}_{\ \ \alpha}$ for which the first term is zero,
and a small perturbation $h a^{\sm(1)}_{\ \ \alpha}$:
\begin{align}
\label{eq:AaPerturbed}
a_\alpha &= a^{\sm(0)}_{\ \ \alpha} + h a^{\sm(1)}_{\ \ \alpha}
.
\end{align}
Then, the Lorenz gauge condition becomes the two conditions:
\begin{align}
\label{eq:LorenzGaugeConditionCovariantPerturbativeDuo0}
\eta^{\alpha\beta} \partial_\alpha a^{\sm(0)}_{\ \ \beta} &=0 \\
\label{eq:LorenzGaugeConditionCovariantPerturbativeDuo1}
h \eta^{\alpha\beta} \partial_\alpha a^{\sm(1)}_{\ \ \beta} &= - h \tilde w e^{\sm(+)\, \alpha\beta}  \partial_\alpha a^{\sm(0)}_{\ \ \beta}
.
\end{align}
Note the resemblance to the form of Equation\ \ref{eq:ScalarWaveEqInhomogForm}. Here, again, the zero-order solution provides a source for the first-order solution.

\subsubsection{Field tensor}\label{sec:FieldTensor}

As noted above, the covariant field tensor is 
\begin{align}
\label{eq:covariantFab}
F_{\mu\nu} &= a_{\nu,\mu} -a_{\mu,\nu}
\end{align}
Because the metric is diagonal and departs from $\eta_{\alpha\beta}$ only in $g_{xx}$ and $g_{yy}$, 
and because those departures are of order $h$ and depend only on $t$ and $z$, it is nearly as easy to form the contravariant field tensor as the covariant field tensor.
The contravariant field tensor is:
\begin{align}
F^{\alpha\beta} 
&= g^{\alpha\mu} g^{\beta\nu}  \left(   \partial_\mu a_\nu - \partial_\nu   a_\mu   \right) \nonumber 
\end{align}
and the Gauss-Ampere law, without charges or currents, is:
\begin{align}
\partial_\beta F^{\alpha\beta} =0 
&= \partial_\beta g^{\alpha\mu} g^{\beta\nu}  \left(   \partial_\mu a_\nu - \partial_\nu   a_\mu   \right)  \nonumber 
\end{align}
For the gravitational-wave spacetime in the coordinates considered here, the metric is diagonal, as is the tensor $e^{\sm(+)\,\alpha\beta}$. Thus, the above two sums collapse to the two indices $\alpha$ and $\beta$.
The expression is nonzero only for the two fixed indices $\alpha=\mu$ and $\beta=\nu$,
so it involves no summations over indices.
This considerably simplifies evaluation.

I collect terms involving $\tilde w$ on the right:
\begin{align}
\partial_\beta g^{\alpha\mu} g^{\beta\nu}  \left(   \partial_\mu a_\nu - \partial_\nu   a_\mu   \right) 
&=\partial_\beta \left( \eta^{\alpha\mu} +h \tilde w e^{\sm(+)\,\alpha\mu} \right)\left(  \eta^{\beta\nu} +h \tilde w e^{\sm(+)\,\beta\nu} \right) \left(   \partial_\mu a_\nu - \partial_\nu   a_\mu   \right) \\
&= \eta^{\alpha\mu} \eta^{\beta\nu} \partial_\beta \left(   \partial_\mu a_\nu - \partial_\nu   a_\mu   \right) \nonumber \\
& + h \Big( 
\left(  e^{\sm(+)\,\alpha\mu} \eta^{\beta\nu} + \eta^{\alpha\mu} e^{\sm(+)\,\beta\nu}\right) \partial_\beta \left(  \tilde w  \left(   \partial_\mu a_\nu - \partial_\nu   a_\mu   \right) \right)
\Big) 
\nonumber
\end{align}
where I discard terms of second order in $h$.

As in Section\ \ref{sec:LorenzGaugeConditionExpand}, I desire a solution that is the sum of a Minkowski-space solution $a^{\sm(0)}_{\ \ \alpha}$,
and a small perturbation $h a^{\sm(1)}_{\ \ \alpha}$:
\begin{align}
a_\alpha &= a^{\sm(0)}_{\ \ \alpha} + h a^{\sm(1)}_{\ \ \alpha}
.
\end{align}
The functions $a^{\sm(0)}_{\ \ \alpha}$ and $h a^{\sm(1)}_{\ \ \alpha}$ must satisfy the equations:
\begin{align}
\label{eq:GaussAmpereLawFundamentalh0}
\eta^{\alpha \mu}  \eta^{\beta \nu} \partial_\beta 
 \left( \partial_\mu a^{\sm(0)}_{\ \ \nu} - \partial_\nu a^{\sm(0)}_{\ \ \mu} \right) 
 &= 0 \\
\label{eq:GaussAmpereLawFundamentalIntermediateh1}
h\, \eta^{\alpha \mu}  \eta^{\beta \nu} \partial_\beta 
\left( \partial_\mu a^{\sm(1)}_{\ \ \nu} - \partial_\nu a^{\sm(1)}_{\ \ \mu} \right)  
 &= - h 
\left(  e^{\sm(+)\,\alpha\mu} \eta^{\beta\nu} + \eta^{\alpha\mu} e^{\sm(+)\,\beta\nu}\right) \partial_\beta \left(  \tilde w  \left(   \partial_\mu a^{\sm(0)}_{\ \ \nu} - \partial_\nu   a^{\sm(0)}_{\ \ \mu}   \right) \right) 
\end{align}
Equation\ \ref{eq:GaussAmpereLawFundamentalh0} is satisfied if $a^{\sm(0)}_{\ \ \nu}$ satisfies the Lorenz gauge condition in Minkowski spacetime, thus canceling the first term in parentheses; and if each component of $a^{\sm(0)}_{\ \ \nu}$ satisfies the scalar wave equation in Minkowski spacetime,
canceling the second term.  The form for $a^{\sm(0)}_{\ \ \nu}$ introduced in Section\ \ref{sec:NearlyPlaneEMWaves} satisfy these conditions by construction.

To simplify Equation\ \ref{eq:GaussAmpereLawFundamentalIntermediateh1}, 
I first differentiate the Lorenz gauge condition, Equation\ \ref{eq:LorenzGaugeConditionCovariantPerturbativeDuo1}, to find:
\begin{align}
\eta^{\alpha\mu} \partial_\mu \left( h\, \eta^{\beta\nu} \partial_\beta a^{\sm(1)}_{\ \ \nu} \right)&= \eta^{\alpha\mu} \partial_\mu \left(- h \tilde w e^{\sm(+)\, \beta\nu}  \partial_\beta a^{\sm(0)}_{\ \ \nu} \right)
\end{align}
I then subtract this from Equation\ \ref{eq:GaussAmpereLawFundamentalIntermediateh1} to find the equation:
\begin{align}
\label{eq:Abstract_GaussAmpereFor_a1_InclRHS}
h\, \eta^{\alpha \mu}  \eta^{\beta \nu} \left( \partial_\beta  \partial_\nu a^{\sm(1)}_{\ \ \mu} \right)  
=&  h 
\left(  e^{\sm(+)\,\alpha\mu} \eta^{\beta\nu} + \eta^{\alpha\mu} e^{\sm(+)\,\beta\nu}\right) \partial_\beta \left(  \tilde w  \left(   \partial_\mu a^{\sm(0)}_{\ \ \nu} - \partial_\nu   a^{\sm(0)}_{\ \ \mu}   \right) \right) \\
&- h \left(  \eta^{\alpha\mu} e^{\sm(+)\,  \beta\nu} \right) \partial_\mu \left( \tilde w  \partial_\beta a^{\sm(0)}_{\ \ \nu} \right) 
\nonumber
\end{align}
The left side of this equation is the Minkowski-spacetime d'Alembertian of $a^{\sm(1)}_{\ \ \nu}$.  The right side is a linear combination of products of $\tilde w$ and $a^{\sm(0)}_{\ \ \nu}$ and their derivatives.
Analogously to the case of the scalar wave equation discussed in Section\ \ref{sec:WaveEqScalarField}, the terms on the right side act as sources for wave equations for each of the 4-vector components of $a^{\sm(1)}$.
Because all the tensors in Equation\ \ref{eq:Abstract_GaussAmpereFor_a1_InclRHS} are diagonal, $\alpha=\mu$ and $\beta=\nu$.  
Equation\ \ref{eq:Abstract_GaussAmpereFor_a1_InclRHS} sums over $\beta=\nu$, but not $\alpha=\mu$.

I now separate the terms involving only derivatives of $a^{\sm(0)}$ from those involving derivatives of $\tilde w$.
I consolidate terms, and use the Gauss-Ampere law and the Lorenz gauge condition for $a^{\sm(0)}$.
I find:
\begin{align}
\label{eq:AbstractSourceTermsForEMWave}
h\, \eta^{\alpha \mu}  \eta^{\beta \nu} \left( \partial_\beta  \partial_\nu \right)  \begin{Bmatrix}  
a^{\sm(1)}_{\ \ t} \\ 
a^{\sm(1)}_{\ \ x} \\ 
a^{\sm(1)}_{\ \ y} \\ 
a^{\sm(1)}_{\ \ z} 
\end{Bmatrix} 
=&  h  \tilde w\, \bigg( \partial_{xx}  -  \partial_{yy} \bigg)
\begin{Bmatrix}  
     -  a^{\sm(0)}_{\ \ t} \\ 
\PM  a^{\sm(0)}_{\ \ x} \\ 
\PM  a^{\sm(0)}_{\ \ y} \\ 
\PM  a^{\sm(0)}_{\ \ z} 
\end{Bmatrix} \\
& +h \left( \partial_t \tilde w \right)
\begin{Bmatrix} 
\PM \partial_x a^{\sm(0)}_{\ \ x} - \partial_y a^{\sm(0)}_{\ \ y}  \\
\PM \partial_x  a^{\sm(0)}_{\ \ t} - \partial_t a^{\sm(0)}_{\ \ x}   \\
     -\partial_y  a^{\sm(0)}_{\ \ t} + \partial_t a^{\sm(0)}_{\ \ y}   \\
0
\end{Bmatrix}
+h \left( \partial_z \tilde w \right)
\begin{Bmatrix} 
0\\
     -  \partial_x  a^{\sm(0)}_{\ \ z} + \partial_z a^{\sm(0)}_{\ \ x}  \\
\PM  \partial_y  a^{\sm(0)}_{\ \ z} - \partial_z a^{\sm(0)}_{\ \ y}  \\
\PM  \partial_x a^{\sm(0)}_{\ \ x} - \partial_y a^{\sm(0)}_{\ \ y} 
\nonumber
\end{Bmatrix}
\end{align}
This expression is quite general; it does not depend on any particular form for $a^{\sm(0)}_{\ \ \alpha}$ or $a^{\sm(1)}_{\ \ \alpha}$, although it does assume the given form for the gravitational wave, and the absence of source charges and currents in the region of validity.

The left side of Equation\ \ref{eq:AbstractSourceTermsForEMWave} is a Minkowski-spacetime scalar wave equation for each 4-vector component of $a^{\sm(1)}_{\ \ \alpha}$.
For each component, the first term on the right side is a ``distributed source'' proportional to undifferentiated $\tilde w$, 
whereas the second and third terms are proportional to derivatives of $\tilde w$. 
The first term involves two derivatives of $a^{\sm(0)}$, and will be proportional to $-k^2\cos[K(-t+z)]$,
for wavelike forms of $a^{\sm(0)}$ with wavenumber $k$.
By contrast, the second and third terms will be proportional to $i kK \sin[K(-t+z)]$. 
Thus, the first term of the distributed source is larger than the rest by a factor of $k/K$.
This is much larger than 1, although I do not ignore the smaller terms here, as outlined in Section\ \ref{sec:Introduction}.

Each 4-vector component $\mu$ of the first term on the right side of Equation\ \ref{eq:AbstractSourceTermsForEMWave}  includes only the same component $\mu$ of $a^{\sm(0)}$.
Moreover, those source terms have the same form as the source terms for the scalar wave equation.
Consequently, if the adopted zero-order solution $a^{\sm(0)}_{\ \ \alpha}$ is a plane wave, 
the two additional scattered waves of the first-order solution can be respresented by the same phase $\tilde \varphi$ found in the scalar-wave case, Equation\ \ref{eq:ScalarVarphidef}.

In contrast to the first term on the right side of Equation\ \ref{eq:AbstractSourceTermsForEMWave},
the second and third terms mix the components of $a^{\sm(0)}_{\ \ \alpha}$.
For example, if $a^{\sm(0)}_{\ \ t}=0$, but $a^{\sm(0)}_{\ \ x}\ne 0$, then $a^{\sm(1)}_{\ \ t}\ne 0$ because $a^{\sm(0)}_{\ \ x}$ contributes to its source in the second term.
If the zero-order solution $a^{\sm(0)}_{\ \ \alpha}$ is a plane wave, 
these terms will introduce two additional scattered waves, as in Equation\ \ref{eq:WaveSum} above.
However, those additional waves cannot be represented as a phase, becuase mixing of the components of the source produces polarization different from the original plane wave.

\section{Nearly-plane electromagnetic waves}\label{sec:NearlyPlaneEMWaves}

\subsection{Introduction: plane waves in Minkowski spacetime}\label{sec:PlaneWavesMinkowski}

I seek plane-wave-like solutions to Maxwell's equations in the spacetime of a gravitational wave,
using Equations\ \ref{eq:LorenzGaugeConditionCovariantPerturbativeDuo0} and\ \ref{eq:GaussAmpereLawFundamentalh0} to find $a^{\sm(0)}$,
and Equations\ \ref{eq:LorenzGaugeConditionCovariantPerturbativeDuo1} and\ \ref{eq:Abstract_GaussAmpereFor_a1_InclRHS} to find $a^{\sm(1)}$.
The following sections take on this task; here I briefly summarize properties of plane electromagnetic waves in Minkowski spacetime,
as an introduction and to provide an expression for $a^{\sm(0)}$.

In Minkowski spacetime, the vector potential for plane wave electromagnetic waves takes the form:
\begin{align}
\label{eq:MinkowskiSpacePlaneWaveForm}
a^{\sm(0)}_{\ \ \alpha} &= A^{\sm(0)}e^{ i k_\beta x^\beta } \enn_\alpha
.
\end{align}
where the physical potential is the real part.
The complex amplitude $A^{\sm(0)}$ gives amplitude and phase.
The covariant electromagnetic wavevector for Minkowski spacetime is:
\begin{align}
\label{eq:kdef}
k_\beta &= \{ k_t, k_x, k_y, k_z \} = \{ -k, k \sin\theta\cos\phi, k \sin\theta\sin\phi, k\cos\theta \}
,
\end{align}
and the contravariant polarization vector is defined as:
\begin{align}
\label{eq:nualdef}
\enn_{\alpha} &= \left( 0, \enn_x, \enn_y, \enn_z \right)
\end{align}
where for convenience l take $n^t=0$, as discussed below. I provide explicit definitions of $\enn_{\alpha}$ in the following section\ \ref{sec:TEandTM}.

The Lorenz gauge condition demands:
\begin{align}
k_\alpha \enn^\alpha = 0
.
\end{align}
If the Lorenz gauge condition holds, 
then Equation\ \ref{eq:GaussAmpereLawFundamentalh0},
the Gauss-Ampere law in Minkowski spacetime in the absence of charges and currents, is satisfied if the 4-wavevector is null: $k_\alpha k^\alpha=0$.

If $\enn^t\ne 0$, a gauge transformation can eliminate $\enn^t$ and thus $a^t$, while
maintaining the form for $a^\alpha$ in Equation\ \ref{eq:MinkowskiSpacePlaneWaveForm} with the same $k_\beta$, and still satisfying the Lorenz gauge condition.
If $\enn^t=0$, then the Lorenz gauge condition simply states that the 3-vector polarization is perpendicular to the spatial components of the wavevector:
\begin{align}
k_j \enn_j =0.
\end{align}
The case $n^t=0$ gives for the electric field $E_j$ and magnetic field $B_j$ of the wave:
\begin{align}
E_j &= i k a_j \\
B_j &= i \epsilon_{j\ell m} k_\ell a_m
\end{align}
where unsubscripted $k=- k_t$ and $\epsilon_{j\ell m}$ is the Levi-Civita symbol.
For simplicity, I will assume $\enn^t=0$ for the Minkowski-space solution, $a^{\sm(0)\, \alpha}$.

\subsection{Polarization}\label{sec:TEandTM}

\begin{figure}
\centering
\includegraphics[width=0.85\textwidth]{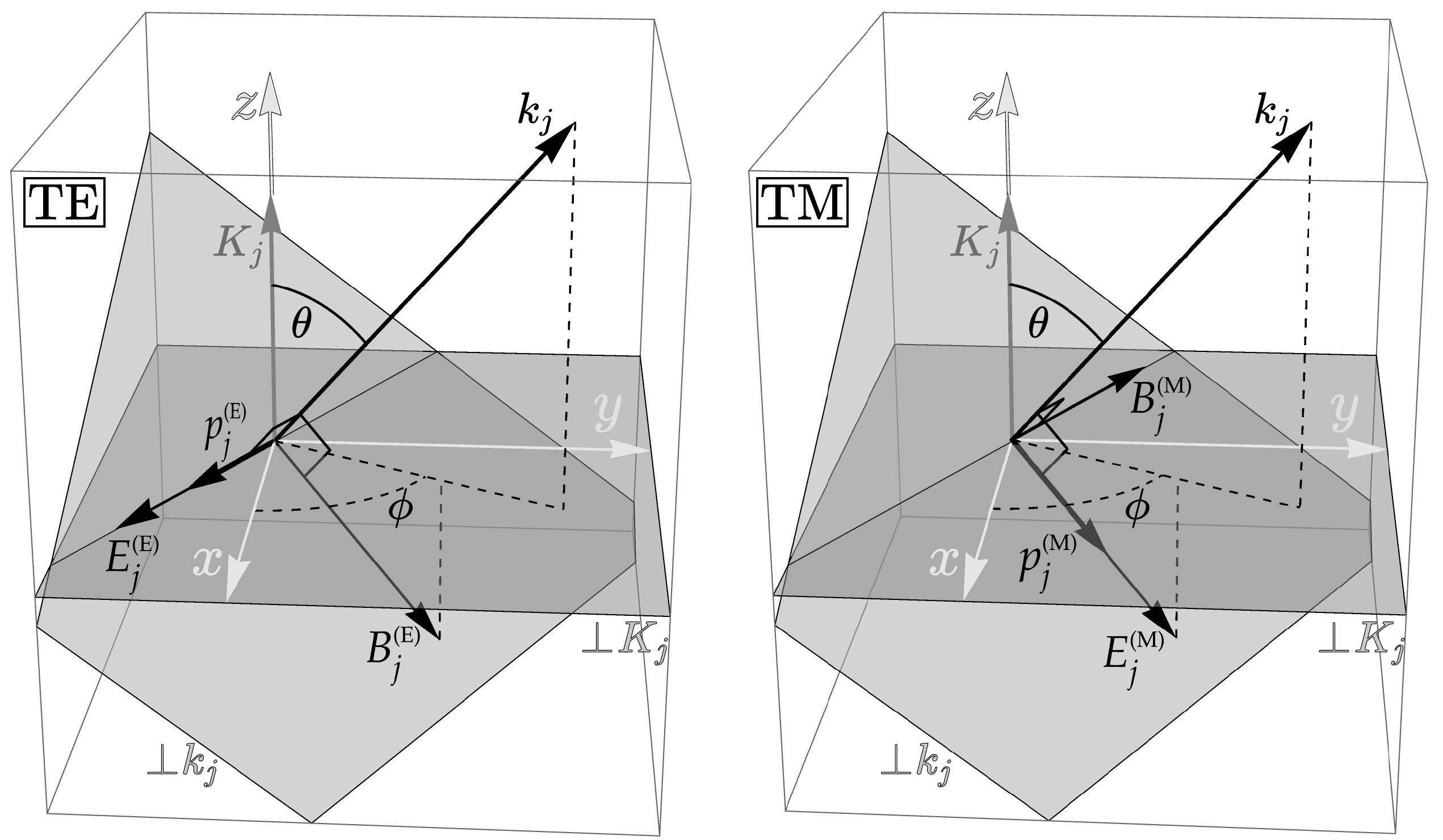} %
\caption{
\label{fig:WaveVectorDiagram}
Spatial geometry for linearly polarized electromagnetic waves.
Left panel: Diagram of 3-vectors for the transverse electric (TE) polarization, in the case $h\rightarrow 0$.
Right panel: Transverse magnetic (TM) polarization.
The fields are shown at the spacetime origin, using the convention of Section\ \ref{sec:TEandTM},
with fields at maximum (positive) amplitude at the instant of these diagrams.  
}
\end{figure}

Plane-wave-like solutions to Maxwell's equations can have many polarization states for a single wavevector $k_\alpha$.
These can be expressed as linear combinations of the two basis states.
As basis states, I choose linear polarizations where, in the limit $h\rightarrow 0$, either the electric field or the magnetic field is perpendicular to both the 3-wavevector of the gravitational wave $K^j$ and the electromagnetic 3-wavevector $k_j$.
The first case is conveniently called transverse electric (TE), and the second transverse magnetic (TM).
The geometries of these polarizations are shown in Figure\ \ref{fig:WaveVectorDiagram}, including wavevectors and other quantities defined below.
I use the superscript $^{\sm(\mathrm{E})}$ to denote the TE case, and $^{\sm(\mathrm{M})}$ the TM case.

One can find the field tensor for the TM polarization from that of the TE polarization using the duality transformation for the electric and magnetic fields (see, for example, \cite{Jackson}):
\begin{align}
\label{eq:DualTensorDefs}
E^{\sm(\mathrm{M})}_{\ \  j} &=\phantom{-}  B^{\sm(\mathrm{E})}_{\ \  j} \\
B^{\sm(\mathrm{M})}_{\ \  j} &= -                  E^{\sm(\mathrm{E})}_{\ \  j}
\nonumber
\end{align}
However, the vector potentials are also interesting, so I will evaluate them separately for both polarizations.

The vector potential for a linearly-polarized TE plane wave, for the Minkowski-space plane-wave given by Equation\ \ref{eq:MinkowskiSpacePlaneWaveForm}, is:
\begin{align}
\label{eq:nualdefTE}
\enn^{\sm(\mathrm{E})}_{\ \ \alpha} &= \left( \enn^{\sm(\mathrm{E})}_t, \enn^{\sm(\mathrm{E})}_x, \enn^{\sm(\mathrm{E})}_y, \enn^{\sm(\mathrm{E})}_z \right) =  \left( 0, \sin\phi, -\cos\phi, 0 \right)
\end{align}
The TM polarization vector is:
\begin{align}
\label{eq:nualdefTM}
\enn^{\sm(\mathrm{M})}_{\ \ \alpha} & = \left( \enn^{\sm(\mathrm{M})}_t, \enn^{\sm(\mathrm{M})}_x, \enn^{\sm(\mathrm{M})}_y, \enn^{\sm(\mathrm{M})}_z \right) = \left( 0, \cos\theta\cos\phi , \cos\theta\sin\phi , -\sin \theta \right)
.
\end{align}
I have taken $n^t=0$ in both cases. Note that the spatial components $\enn^{\sm(\mathrm{E})}_{\ \ j}$ and $\enn^{\sm(\mathrm{M})}_{\ \ j}$ are perpendicular, as expected from the fact that $\bm{E}\perp\bm{B}$ and the duality transformation,
Equation\ \ref{eq:DualTensorDefs}.

\subsection{Candidate solution}\label{sec:CandidateSolution}

I seek a solution of the Lorentz gauge condition and the the Gauss-Ampere law that takes a plane-wave form in the limit $h\rightarrow 0$.
The solution takes the form (Equation\ \ref{eq:AaPerturbed}):
\begin{align}
a_\alpha &= a^{\sm(0)}_{\ \ \alpha} + h a^{\sm(1)}_{\ \ \alpha}
\end{align}
where $a^{\sm(0)}_{\ \ \alpha}$ is given by Equation\ \ref{eq:MinkowskiSpacePlaneWaveForm}.
This form satisfies the zero-order Lorenz gauge condition, 
Equation\ \ref{eq:LorenzGaugeConditionCovariantPerturbativeDuo0},
and the Gauss-Ampere law, Equation\ \ref{eq:GaussAmpereLawFundamentalh0}.

I thus seek $h a^{\sm(1)}_{\ \ \alpha}$ that satisfies the first-order Lorenz gauge condition, Equation\ \ref{eq:LorenzGaugeConditionCovariantPerturbativeDuo1},
and the first-order Gauss-Ampere law, Equation\ \ref{eq:AbstractSourceTermsForEMWave}.
These equations are analogous to the first-order scalar wave equation:
the left side is the form appropriate for Minkowski spacetime, but involves $a^{\sm(1)}_{\ \ \alpha}$ rather than $a^{\sm(0)}_{\ \ \alpha}$;
and the right side involves products of derivatives of $a^{\sm(0)}_{\ \ \alpha}$ and $\tilde w$ and its derivatives.
The space and time dependences of the terms on the right are weighted sums of the additional waves found for the scalar wave equation in Section\ \ref{sec:WaveEqScalarField}, $\exp[ i (k_\alpha + K_\alpha)x^\alpha]$ and $\exp[ i (k_\alpha - K_\alpha)x^\alpha]$,
or equivalently weighted sums of $\cos[K(-t+z)]e^{i k_\mu x^\mu}$ and $\sin[K(-t+z)]e^{i k_\mu x^\mu}$.
The left sides are linear, 
and $\exp[ i (k_\alpha + K_\alpha)x^\alpha]$ and $\exp[ i (k_\alpha - K_\alpha)x^\alpha]$ are eigenfunctions of the left-side operators, 
so we expect the first-order solution to be a superposition of these waves.

In terms of the discussion of observable delay and direction in Section\ \ref{sec:PhaseDelayDirectionIntro},
and as found for the scalar wave equation,
the right sides of Equations\ \ref{eq:LorenzGaugeConditionCovariantPerturbativeDuo1} and\ \ref{eq:AbstractSourceTermsForEMWave} will introduce additional, scattered waves.
As argued immediately following Equation\ \ref{eq:AbstractSourceTermsForEMWave}, the phase $\tilde \varphi$ satisfies the first term on the right side of that equation,
just as for the scalar wave equation.
This corresponds to two additional waves.
However, it cannot satisfy the Lorenz gauge condition, or the remaining two terms on the right side of Equation\ \ref{eq:LorenzGaugeConditionCovariantPerturbativeDuo1},
because these equations mix the Cartesian components.
For example, if $a^{\sm(0)}_{\ \ t}=0$ and $a^{\sm(0)}_{\ \ x}\ne 0$,
then $\partial_t a^{\sm(1)}_{\ \ t} \ne 0$ in the first-order Lorenz gauge condition, Equation\ \ref{eq:LorenzForCandidateCondition1}.
The same holds for the second two terms on the right of the Gauss-Ampere law.
Thus, $a^{\sm(1)}_{\ \ \alpha}$ will have a different polarization from $a^{\sm(0)}_{\ \ \alpha}$.
A phase variation cannot produce polarization variations, so $a^{\sm(1)}_{\ \ \alpha}$ will involve at least two additional waves,
for a total of at five waves that combine to define delay and direction.

As a candidate nearly-plane wave solution for an electromagnetic wave in gravitational-wave spacetime as described in Section\ \ref{sec:GravWaveDefs}, I investigate the vector potential:
\begin{align}
\label{eq:aCandidateVectorPotentialDef}
a_{\alpha}& = A^{\sm(0)} \exp \left[ i k_\beta x^\beta + i h \tilde\varphi \right] \left( \enn_\alpha + h \tilde f  q_\alpha \right) \\
&= a^{\sm(0)\, \alpha} + h a^{\sm(1)\, \alpha} + \mathcal{O}\left[h^2\right] \nonumber \\
&= a^{\sm(0)\, \alpha} + i h \tilde \varphi  a^{\sm(0)\, \alpha} + h \tilde f  q_\alpha  A^{\sm(0)} \left[ i k_\beta x^\beta \right]  + \mathcal{O}\left[h^2\right]
,
\nonumber
\end{align}
where $A^{\sm(0)}$ is a complex constant. 
The physical fields are the real part of $a^{\alpha}$.
I propose this form as a solution for Equation\ \ref{eq:AbstractSourceTermsForEMWave}.

The phase $\tilde\varphi$ turns out to be the same as for the scalar wave equation, as discussed in the text following Equation\ \ref{eq:AbstractSourceTermsForEMWave}.
I will calculate $\tilde \varphi$ independently below, and demonstrate that the two are indeed equal.
As discussed above, a phase variation cannot account for the polarization variations implied by the Lorenz gauge condition and the Gauss-Ampere law.
Consequently, the candidate solution
requires the additional function $\tilde f$ and polarization vector $q_\alpha$, which leads to amplitude or polarization changes.
I will assume that both $\tilde \varphi$ and $\tilde f$ are real functions of space and time, 
and that they depend only on $t$ and $z$, because the problem is independent of $x$ and $y$.

The gravitational-wave-induced polarization $q_\alpha$ is a 4-vector constant with space and time, but may depend on the parameters $k_\alpha$ and $\enn^\alpha$:
\begin{align}
q_\alpha &= \left\{ q_t, q_x, q_y, q_z \right\}
.
\end{align}
Because $q_\alpha$ depends on $\enn_\alpha$, as discussed below, I will adopt the notations
$q^{\sm(\mathrm{E})}_{\ \ \alpha}$ and $q^{\sm(\mathrm{M})}_{\ \ \alpha}$ for transverse electric and transverse magnetic polarization, respectively, when evaluating the fields in these cases.

The zero-order candidate is the plane-wave solution of Section\ \ref{sec:PlaneWavesMinkowski}:
\begin{align}
\label{eq:a0Lorenzh0}
a^{\sm(0)}_{\ \ \alpha} &= A^{\sm(0)} e^{ i k_\beta x^\beta }  \enn_\alpha
\end{align}
and the first-order candidate is
\begin{align}
\label{eq:a0Lorenzh1}
h a^{\sm(1)}_{\ \ \alpha} &= h A^{\sm(0)} e^{ i k_\beta x^\beta }  \left( i\tilde\varphi \enn_\alpha + \tilde f q_\alpha \right)
.
\end{align}

\subsubsection{Lorenz gauge condition for candidate Solution}\label{sec:LorenzForCandidate} 

The candidate solution in Equation\ \ref{eq:aCandidateVectorPotentialDef} 
must satisfy the Lorenz gauge condition given by Equation\ \ref{eq:LorenzGaugeConditionCovariant}.
In particular, 
the zero-order candidate $a^{\sm(0)}_{\ \ \alpha}$ must satisfy the Lorenz gauge condition for $h= 0$, Equation\ \ref{eq:LorenzGaugeConditionCovariantPerturbativeDuo0}, so that:
\begin{align}
 k_\alpha \eta^{\alpha\beta} \enn_\beta=0
\end{align}
as expected.
The first-order candidate $a^{\sm(1)}_{\ \ \alpha}$ 
must satisfy the corresponding Lorenz gauge condition Equation\ \ref{eq:LorenzGaugeConditionCovariantPerturbativeDuo1}:
\begin{align}
\label{eq:TELorenzTildef}
h \left( \partial_\alpha \tilde f \right) \eta^{\alpha\beta} q_\beta + 
i h \left( k_\alpha   \eta^{\alpha\beta} q_\beta \tilde f  + \left( \partial_\alpha \tilde\varphi \right) \eta^{\alpha\beta} \enn_\beta  \right) &= - i h k_\alpha  \tilde w e^{\sm(+)\, \alpha\beta} \enn_\beta 
.
\end{align}
This equation can be satisfied if $\tilde f$ is complex, contrary to my assumption above; in that case, $\tilde f$ takes on part of the role of $\tilde\varphi$.
However, it is simpler to suppose that $\tilde f$ is real, and regard the real and imaginary parts of Equation\ \ref{eq:TELorenzTildef} as two independent conditions:
\begin{align}
\label{eq:LorenzForCandidateCondition0}
h \left( \partial_\alpha \tilde f \right) \eta^{\alpha\beta} q_\beta &= 0  \\
\label{eq:LorenzForCandidateCondition1}
i h \left( k_\alpha   \eta^{\alpha\beta} q_\beta \tilde f  + \left( \partial_\alpha \tilde\varphi \right) \eta^{\alpha\beta} \enn_\beta  \right) &= - i h k_\alpha \tilde w e^{\sm(+)\, \alpha\beta} \enn_\beta
.
\end{align}
I argued above that $\tilde f$ depends only on $t$ and $z$. I will argue below that $\partial_t \tilde f = - \partial_z \tilde f$, as is the case for $\tilde w$.
In this case, Equation\ \ref{eq:LorenzForCandidateCondition0} implies:
\begin{align}
\label{eq:qdt=-qdz}
q_t &= - q_z
\end{align}

\subsubsection{Gauss-Ampere law and solution for parameters of candidate solution}\label{sec:GAandSolutionForCandidate}

As well as the Lorenz gauge condition, the candidate solution in Equation\ \ref{eq:aCandidateVectorPotentialDef} 
must satisfy the Gauss-Ampere law.
Equations\ \ref{eq:GaussAmpereLawFundamentalh0} and\ \ref{eq:AbstractSourceTermsForEMWave}
give equivalent forms of this law, if the Lorenz gauge condition holds.
In this work, I use Equation \ref{eq:AbstractSourceTermsForEMWave} because it has convienient properties, as discussed following that equation above.

Terms on the right side of Equation\ \ref{eq:AbstractSourceTermsForEMWave} involving undifferentiated $\tilde w$ and derivatives of $\tilde w$ are real and imaginary, respectively,
because the candidate solution represents the electromagnetic wave as a complex exponential: each additional derivative introduces a factor of $i$.
This provides a simple way of displaying them separately:
\begin{alignat}{5}
\label{eq:Abstract_GaussAmpereFor_a1_Evaluate}
\scriptstyle{t  :\,\mathrm{Re :}}\ \ &&                                                                                                              0  &=& 0  \nonumber \\
\scriptstyle{x :\,\mathrm{Re :}}\ \ &&  2 h\,  (\enn_x ( k_t(\partial_t \tilde \varphi) -k_z (\partial_z \tilde \varphi) ) )  &=&  h\, \Big(  &&    (-k_x^2 + k_y^2) \enn_x   &\cos[K(-t+z)]    \Big)  \\ 
\scriptstyle{y :\,\mathrm{Re :}}\ \ &&  2 h\,  (\enn_y ( k_t(\partial_t \tilde \varphi) -k_z (\partial_z \tilde \varphi) ) )  &=&  h\, \Big(  &&    (-k_x^2 + k_y^2) \enn_y   &\cos[K(-t+z)]   \Big) \nonumber \\
\scriptstyle{z :\,\mathrm{Re :}}\ \ &&  2 h\,  (\enn_z ( k_t(\partial_t \tilde \varphi) -k_z (\partial_z \tilde \varphi) ) )  &=&  h\, \Big( &&    (-k_x^2 + k_y^2) \enn_z   &\cos[K(-t+z)]    \Big) \nonumber \\
 \nonumber \\
\scriptstyle{t :\,\mathrm{Im :}}\ \ &&  2 i h ( q_t (k_t (\partial_t \tilde f) -k_z (\partial_z \tilde f ))) &=&   i h K \Big(&&-(k_x \enn_x - k_y \enn_y ) &\sin[K(-t+z)]   \Big)\nonumber \\
\scriptstyle{x :\,\mathrm{Im :}}\ \ &&  2 i h ( q_x (k_t (\partial_t \tilde f) -k_z (\partial_z \tilde f ))) &=&   i h K \Big(&&-(k_t + k_z) \enn_x + k_x \enn_z) &\sin[K(-t+z)]   \Big) \nonumber \\
\scriptstyle{y :\,\mathrm{Im :}}\ \ &&  2 i h ( q_y (k_t (\partial_t \tilde f) -k_z (\partial_z \tilde f ))) &=&   i h K \Big(&&(k_t + k_z) \enn_y - k_y \enn_z) &\sin[K(-t+z)]   \Big) \nonumber \\
\scriptstyle{z :\,\mathrm{Im :}}\ \ &&  2 i h ( q_z (k_t (\partial_t \tilde f) -k_z (\partial_z \tilde f ))) &=&  i h K \Big(&&(-k_x \enn_x + k_y \enn_y ) &\sin[K(-t+z)]   \Big) \nonumber 
\end{alignat}
Here, I assume that $\partial_{tt} \tilde \varphi = \partial_{zz} \tilde\varphi$ and $\partial_{tt} \tilde f = \partial_{zz} \tilde f$, because they are linear combinations of $\cos[K(-t+z)]$, and $\sin[K(-t+z)]$.
This equation takes precisely the form expected from the discussion following Equation\ \ref{eq:AbstractSourceTermsForEMWave}.
The first of the three terms in the distributed source lead to the real terms, with coefficient $\cos[K(-t+z)]$; the second and third terms lead to the imaginary terms with coefficient $\sin[K(-t+z)]$.
The function $\tilde \varphi$ depends on the real terms, and $\tilde f$ on the imaginary.
Both functions are real, 
but $h \tilde \varphi$ is multiplied by $i$ in the candidate solution, Equation\ \ref{eq:aCandidateVectorPotentialDef}.
Therefore,
the factor of $i h \tilde \varphi$ leads to a phase change, and $h \tilde f q_\alpha$ to a change in amplitude or polarization.
The 3-vector components of $\tilde f a_\alpha$ that are parallel to $\enn_\alpha$ lead to an amplitude change, and the perpendicular ones lead to a polarization change.
The time components of $\tilde f a_\alpha$ can lead to either.
The description of $a^{\sm(1)}_{\ \ \alpha}$ thus requires at least four scattered waves.
The two waves corresponding the $\tilde \varphi$ can be combined into a single, global phase.

Inspection of the real parts of Equation\ \ref{eq:Abstract_GaussAmpereFor_a1_Evaluate} 
shows that $ \tilde \varphi$ is proportional to $\sin[K(-t+z)]$, and $p_\alpha$ cancels.
Then, by precisely the argument used for the scalar wave equation in Section\ \ref{sec:WaveEqScalarField},
\begin{align}
\label{eq:EMtildevarphiDef}
h \tilde \varphi &= -\frac{h}{2} \frac{k}{K} (1+\cos\theta) \cos[2\phi] \sin[K(-t+z)]
\end{align}
for all nonzero entries. This is the same as the result for $\tilde \varphi$ for the scalar wave equation, and for the delay for the geodesic calculation.

Inspection of the imaginary part shows that $\tilde f$ is proportional to $\cos[K(-t+z)]$ and that $q_\alpha$ depends on $\enn_\alpha$.
I take:
\begin{align}
\label{eq:EMtildefDef}
h \tilde f[t,z] &=h \cos[K(-t+z)]
\end{align}
Evaluation with the two polarizations yields:
\begin{align}
\label{eq:qdalphaTEdef}
q^{\sm(E)}_{\ \, \alpha} &=
\left\{  \frac{\sin\theta\sin[2\phi]}{2(1-\cos\theta)} , \haf \sin\phi , \haf \cos\phi , -\frac{\sin\theta\sin[2\phi]}{2(1-\cos\theta)}  \right\}  \\
\label{eq:qdalphaTMdef}
q^{\sm(M)}_{\ \, \alpha} &=
\left\{ \frac{\cos\theta\sin\theta\cos[2\phi]}{2(1-\cos\theta)} , -\haf \cos\phi , \haf \sin\phi , -\frac{\cos\theta\sin\theta\cos[2\phi]}{2(1-\cos\theta)}  \right\}
\end{align}
Interestingly, the $t$ and $z$ components of $q_\alpha$ are singular at $\theta=0$, for both polarizations. 
These singularities do not appear in the fields, as I demonstrate below.
Note that the solutions above satisfy the Lorenz gauge conditions, Equations\ \ref{eq:LorenzForCandidateCondition0} and\ \ref{eq:LorenzForCandidateCondition1}, and the Gauss-Ampere law, Equations\ \ref{eq:GaussAmpereLawFundamentalh0} and\ \ref{eq:AbstractSourceTermsForEMWave}.
Calculation of the fields is straightforward, but I will present the fields in the local Lorentz frame of the observer, in Section\ \ref{sec:FieldsLorentzFrame} below.

\section{Observed delay and direction: scalar field}\label{sec:DelayDirectionScalar}

An observer can measure the phase, delay, direction, intensity, and polarization of the wave. These quantities are evaluated in the local Lorentz frame of the observer. 
The local Lorentz frame extends over a limited time, over a limited spatial region, about the event of observation.
The dimensions of that region, in space and time, are approximately $1/K$.

For the scalar wave discussed in Section\ \ref{sec:WaveEqScalarField},
the gravitational wave introduces two additional scattered waves.
However, the effects of those waves can be equally well represented by the phase change $h \tilde \varphi$. 
The observed delay and direction are calculable from $\varphi$,
as outlined in Section\ \ref{sec:PhaseDelayDirectionIntro}.

\subsection{Observed phase and delay}

The observed phase is a scalar; indeed, it is (in principle) measurable by the observer,
for example if the source transmits phase-locked time signals.
Thus, the phase at an event must be the same in all frames.
The observed phase at $x^\alpha$ is simply:
\begin{align}
\Phi [x^\alpha] &= k_\beta x^\beta + h \tilde \varphi (x^\alpha) 
.
\end{align}
The delay is given by the derivative of phase with respect to wavenumber, evaluated at the event of observation $x^{{\sm( \mathrm o )}\, \alpha}$:
\begin{align}
\tau &=  \left[ \partial_{k} \left(   k_\alpha x^\alpha + h \tilde \varphi \right) \right]_{x^{{\sm( \mathrm o )}\, \alpha}} 
\end{align}
For convenience, I will assume that the observer is at the spatial origin, and observes at time $t^{\sm( \mathrm o )}$.
Then,
\begin{align}
\tau &= t^{\sm( \mathrm o )} - h \frac{1}{2 K} (1+\cos \theta) \cos[2\phi] \sin[ K t^{\sm( \mathrm o )}]
.
\end{align}
The first term is the result for $h=0$; I simply ignore it here.
The second term is the change in delay caused by the gravitational wave.
The result here is the same as those previously found in the geometric-optics limit \cite{EstabrookWahlquist1975GReGr...6..439E,Detweiler1979ApJ...234.1100D,Pyne1996,BookFlanagan11}
and by Kopeikin et al.  \cite{Kopeikin99,1999PhRvD..60l4002K,2014frc1.book.....K} in the plane-wave case. The delay has order of magnitude $h/K$, and has dimensions of time.
It is quite small for gravitational waves with $h\sim 10^{-20}$ and $K\lesssim 10^{-3}$.
The corresponding phase has order of magnitude $h k/K$ and is dimensionless.  It can be of order 1, or even greater;
but measurement would require knowledge of the phase at the source.
Such a measurement might be possible for a phase-locked clock on a spacecraft at distance beyond a gravitational-wave wavelength, or a few AU.

\subsection{Apparent direction}\label{sec:ApparentSourceDirection}

The apparent direction of a source at an observer is a unit vector opposite to the direction of the spatial gradient of the wave phase, in the local Lorentz frame.
For example, for the waves discussed in this paper, in the limit $h\rightarrow 0$, the source direction is opposite the spatial wavevector $k^j$. 
Only an instrument extended in space, such as an interferometer array or a lens with finite diameter,
can measure the wave phase over a region of spacetime, and thus find the gradient and determine the source direction.
Direction is thus more subtle than delay, which can be measured at a single spatial position, although it does require an interval of time greater than $2\pi/k$..

I suppose that the observer measures source direction at the spatial origin, at coordinate time $t^{\sm( \mathrm o )}$.
Pyne et al. (\cite{Pyne1996}, Equation 40)
define orthogonal Cartesian coordinates in the local Lorentz frame near the origin $x^{\alpha'}$, by normalizing $x$ and $y$:
\begin{align}
\label{eq:OrthoCartesianLocalLorentzFrame}
t' &= t - t^{\sm( \mathrm o )}\\
x' &= (1+ \haf h \tilde w^{\sm( \mathrm o )}) x \nonumber \\
y' &= (1-  \haf h \tilde w^{\sm( \mathrm o )}) y \nonumber \\
z' &= z \nonumber 
,
\end{align}
where
\begin{align}
\tilde w^{\sm( \mathrm o )}&= \cos[K t^{\sm( \mathrm o )}]
.
\end{align}
The primed frame is the local Lorentz frame, near the observer at $|t'|,\ |x'_j| \ll 1/ K$.
This transformation holds not only for events near the observer, but for any contravariant vector in that region.

The observed source direction $\hat s^{j'}$ is a unit 3-vector, opposite the direction of the spatial gradient of phase as measured in the observer's local Lorentz frame.
The 4-vector 
$\left. \partial^{\alpha'} \Phi \right|_{\{t'^{\sm(\mathrm{o})}, 0\}}$
is null for a wave that is locally approximately a plane wave. 
Hence, I normalize the spatial components that comprise the source direction by the time component:
\begin{align}
\hat s^{j'} &= - \frac{1}{\left| \left. \partial^{t'} \Phi \right|_{\{t'^{\sm(\mathrm{o})}, 0\}} \right| }   \left. \partial_{j'} \Phi \right|_{\{t'^{\sm(\mathrm{o})}, 0\}}
,
\end{align}
where the sign indicates that the direction is opposite to the phase gradient.
For example, for $a^{\sm(0)}$ in the candidate solution, $\left. \partial^{\alpha'} \Phi \right|_{\{t'^{\sm(\mathrm{o})}, 0\}}\rightarrow k^\alpha$,
and the source direction is $-k^j/|k^t|$.

For convenience, I compute the derivatives of phase $\Phi$ in the coordinate frame,
and then transform to the observer's local Lorentz frame using Equation\ \ref{eq:OrthoCartesianLocalLorentzFrame}.
I find:
\begin{alignat}{3}
\label{eq:PhaseInCartesianLocalLorentzFrame}
\partial_{t'} \Phi &= -\partial^{t'} \Phi &&= -\partial^{t} \Phi                                   &&= \partial_{t} \Phi \\
\partial_{x'} \Phi &=  \partial^{x'} \Phi &&= (1+ \haf h \tilde w)  \partial^{x} \Phi  &&=(1+ \haf h \tilde w) (1- h \tilde w)  \partial_{x} \Phi  \nonumber \\
\partial_{y'} \Phi &=  \partial^{y'} \Phi &&= (1- \haf h \tilde w)  \partial^{y} \Phi  &&= (1- \haf h \tilde w)  (1+ h \tilde w)  \partial_{y} \Phi  \nonumber \\
\partial_{z'} \Phi &=  \partial^{z'} \Phi &&=  \partial^{z} \Phi                              &&=  \partial_{z} \Phi \nonumber 
.
\end{alignat}

For the gravitational-wave spacetime, the phase is (Equation\ \ref{eq:EMtildevarphiDef}):
\begin{align}
\Phi 
&= k_\alpha x^\alpha - \frac{h}{2} \frac{k}{K} (1+\cos\theta) \cos[2\phi] \sin[K(-t+z)]  
, 
\end{align}
and its derivatives are, through first order in $h$:
\begin{align}
\frac{1}{\left| \left. \partial_t \Phi \right|_{\{t^{\sm( \mathrm o )},0\}} \right| } &= \left( \left| -k + \haf h k (1+\cos\theta)\cos[2 \phi] \cos[K t^{\sm( \mathrm o )}] \right| \right)^{-1}   \approx \frac{1}{k}\left( 1 + \haf h  (1+\cos\theta)\cos[2\phi] \cos[K t^{\sm( \mathrm o )}] \right) \\
\left. \partial_j \Phi \right|_{\{t^{\sm( \mathrm o )},0\}} &= k_j -\haf h k  (1+\cos\theta)\cos[2\phi] \cos[K t^{\sm( \mathrm o )}] \delta_{j z} \nonumber
\end{align}

Consequently, using Equation\ \ref{eq:PhaseInCartesianLocalLorentzFrame} I find:
\begin{align}
\hat s^{x'} &=-  \left( 1 + \haf h  (1+\cos\theta)\cos[2\phi] \cos[K t^{\sm( \mathrm o )}] \right)  (1- \haf h \cos[K t^{\sm( \mathrm o )}] ) \sin\theta \cos\phi    \\
\hat s^{y'} &=-  \left( 1 + \haf h  (1+\cos\theta)\cos[2\phi] \cos[K t^{\sm( \mathrm o )}] \right)  (1+ \haf h \cos[K t^{\sm( \mathrm o )}] ) \sin\theta \sin\phi \nonumber \\
\hat s^{z'} &=-  \left( 1 + \haf h  (1+\cos\theta)\cos[2\phi] \cos[K t^{\sm( \mathrm o )}] \right) ( \cos\theta -\haf h (1+\cos\theta)\cos[2\phi] \cos[K t^{\sm( \mathrm o )}] ) 
.
\nonumber
\end{align}
The parameters $\{\theta,\phi\}$ indicate the electromagnetic wave direction.
However, the source is in the opposite direction, at a distance approaching infinity for a perfect plane wave, as assumed here.
An observer may 
prefer to express the position offset due to the gravitational wave in terms of the angular position of the source $\{\theta^{\sm(\mathrm{s})}, \phi^{\sm(\mathrm{s})}\}$,
rather than in terms of the wavevector $\{\theta,\phi\}$.  
These are related by:
\begin{align}
\theta^{\sm(\mathrm{s})} & = \pi - \theta \\
\phi^{\sm(\mathrm{s})} &=\phi + \pi \nonumber
\end{align}
In terms of the unit vectors of the observer-centered spherical coordinates (see \cite{Arfken}, Problem 3.10.19)
the source direction is:
\begin{align}
\bm{ \hat s }' &= -\mathbf{\hat k} + \haf h \sin\theta^{\sm(\mathrm{s})} \left( \cos\left[2\phi^{\sm(\mathrm{s})}\right] {\bm{ \hat \theta }^\sm(\mathrm{s})} + \sin\left[2\phi^{\sm(\mathrm{s})}\right] {\bm{ \hat \phi }^{\sm(\mathrm{s})}} \right) \cos \left[ K t^{\sm( \mathrm o )} \right] 
,
\end{align}
where $\mathbf{\hat k}=\{\sin\theta\cos\phi,\sin\theta\sin\phi,\cos\theta\}= -\{ \sin\theta^{\sm(\mathrm{s})}\cos\phi^{\sm(\mathrm{s})},\sin\theta^{\sm(\mathrm{s})}\sin\phi^{\sm(\mathrm{s})},\cos\theta^{\sm(\mathrm{s})} \}$.
The derivative of $ \bm{ \hat s }'$ with respect to $t^{\sm( \mathrm o )}$ is the observed proper motion of the source. 
These results agree with those of Pyne et al. \cite{Pyne1996} (where the direction of $\theta=0$ is opposite to that used here), and with \citet{Kopeikin99} in the plane-wave case.
They agree with Book and Flanagan \citep{BookFlanagan11} in the local Lorentz frame.
 
\section{Observed delay, polarization, direction, and intensity: electromagnetic field}\label{sec:DelayDirectionEM}

For the electromagnetic wave, the effects of the gravitational wave are represented by the same phase change $h \tilde \varphi$,
as well as a change in the fields through the contribution of $q_\alpha$ and $\tilde f$ to $h a^{\sm(1)}_{\ \ \alpha}$.
This can be expressed as the creation of four additional scattered waves. Two are the same as those for the scalar field.
Two others vary through spacetime with the same sum and difference 4-wavevectors, $(k_\alpha+K_\alpha)$ and $(k_\alpha-K_\alpha)$,
but have different amplitude and polarization.
The resulting field still has the overall phase $\Phi = k_\alpha x^\alpha +h \tilde \varphi[t,z]$ in the neighborhood of the observer,
and from this the observer calculates the same delay $\tilde \varphi$ as in the scalar wave case.

On the other hand, calculation of the apparent direction is complicated by need to superpose all five waves.
In principle, the difference among the wavevectors is small enough that the observer can superpose the five waves within a region with dimensions much smaller than $1/K$,
to produce one wave in the neighborhood of the observer, with some net wavevector, net polarization, and net phase offset.
However, in this case the apparent direction is most easily calculated from the Poynting 3-vector, as noted in Section\ \ref{sec:PhaseDelayDirectionIntro}.
I will calculate the fields, the Poynting vector and the direction in the following sections.
 
\subsection{Observed electromagnetic fields}\label{sec:FieldsLorentzFrame}

The electric and magnetic fields can be easily calculated from the results presented in Section\ \ref{sec:GAandSolutionForCandidate}.
An observer measures these fields in his local Lorentz frame.
Conversion of the covariant field tensor to this frame requires the transformation for covariant quantities:
 \begin{align}
 \label{eq:GNSigmaFrameTransform}
u_{\alpha'} &= \left( I^{\ \ \beta}_{\alpha'} - \haf h \tilde w e^{\sm(+)\, \beta}_{\alpha'}\right) u_\beta
\end{align}
where where $I^{\ \ \beta}_{\alpha'}  = I^{\alpha'}_{\ \ \beta} $ is the identity matrix, and 
\begin{align}
e^{\sm(+)\, \beta}_{\alpha'} &= e^{\sm(+)\, \alpha'}_{\ \ \ \ \, \beta} =
\begin{cases}
+1 & \alpha'=\beta=x \\
-1 & \alpha'=\beta=y \\
0 & \mathrm{otherwise.}
\end{cases}
.
\end{align}
By contrast, Equation\ \ref{eq:OrthoCartesianLocalLorentzFrame} above expresses the transformation for contravariant quantities:
\begin{align}
x^{\alpha'} &= \left( I^{\alpha'}_{\ \ \beta} + \haf h \tilde w e^{\sm(+)\, \alpha'}_{\ \ \ \ \, \beta}\right) x^\beta
.
\end{align}
Note that the symbols $I^{\ \ \beta}_{\alpha'},\  I^{\alpha'}_{\ \ \beta},\ e^{\sm(+)\, \beta}_{\alpha'},\ e^{\sm(+)\, \alpha'}_{\ \ \ \ \, \beta} $ are not 
covariant tensors; they are valid only in the observer's local Lorentz frame, and indices cannot be raised or lowered.

Calculation of the field tensor with the evaluated candidate solution is straightforward,
using Equation\ \ref{eq:covariantFab} for the field tensor and the definition of the candidate solution in Equation\ \ref{eq:aCandidateVectorPotentialDef},
and $k_\alpha$ from Equation\ \ref{eq:kdef}.
I substitute for $\tilde \varphi$ using Equation\ \ref{eq:EMtildevarphiDef}, for $\tilde f$ using Equation\ \ref{eq:EMtildefDef}. For the TE mode, I use $n_{\ \ \alpha}^{\sm(\mathrm{E})}$  as defined in Equation\ \ref{eq:nualdefTE} and $q_{\ \ \alpha}^{\sm(\mathrm{E})}$ from Equation\ \ref{eq:qdalphaTEdef}.

I then transform to the local Lorentz frame of the observer using Equation\ \ref{eq:GNSigmaFrameTransform}:
\begin{align}
F_{\alpha' \beta'} &=\left(I^{\alpha\alpha'}  - \haf h \tilde w e^{\sm(+)\, \alpha}_{\alpha'} \right)\left( I^{\beta\beta'} - \haf h \tilde w e^{\sm(+)\, \beta}_{\beta'} \right) F_{\alpha \beta} 
\end{align}
This yields the components of $F^{\sm(\mathrm{E})}_{\ \ \alpha'\beta'}$, to first order in $h$:
\begin{alignat}{2}
\label{eq:TE_symbolic}
E^{\sm(\mathrm{E})}_{\ \  j'}  &=  
 \begin{Bmatrix} F^{\sm(\mathrm{E})}_{\ \  t' x'}\\F^{\sm(\mathrm{E})}_{\ \  t' y'}\\F^{\sm(\mathrm{E})}_{\ \  t' z'} \end{Bmatrix} =
- \begin{Bmatrix} F^{\sm(\mathrm{E})}_{\ \  x' t'}\\F^{\sm(\mathrm{E})}_{\ \  y' t'}\\F^{\sm(\mathrm{E})}_{\ \  z' t'} \end{Bmatrix} 
\\
&= A^{\sm(0)} k e^{ i k_\beta x'^\beta+ih\tilde\varphi}  \left(                                         \    \   i k                                               \begin{Bmatrix} -\sin\phi \\ \cos\phi \\ 0 \end{Bmatrix}   \right. 
                                                                                                                              &           - \frac{i}{2}  h k \cos[K(-t+z)]     \begin{Bmatrix} (1+\cos\theta)\sin\phi\\ (1+\cos\theta)\cos\phi\\ -\sin\theta \sin[2\phi] \end{Bmatrix} \quad\ 
                                                                                                    \nonumber \\&   &\left.     + \frac{1}{2} h K \sin[K(-t+z)]     \begin{Bmatrix} \sin\phi\\  \cos\phi\\ 0 \end{Bmatrix}    \right)
                                                                                                   ,\nonumber\\
B^{\sm(\mathrm{E})}_{\ \  j'} &=
 \begin{Bmatrix} F^{\sm(\mathrm{E})}_{\ \  y' z'}\\ -F^{\sm(\mathrm{E})}_{\ \ x' z'}\\F^{\sm(\mathrm{E})}_{\ \  x' y'} \end{Bmatrix} = \begin{Bmatrix} -F^{\sm(\mathrm{E})}_{\ \  z' y'}\\  F^{\sm(\mathrm{E})}_{\ \  z' x'}\\ -F^{\sm(\mathrm{E})}_{\ \  y' x'} \end{Bmatrix}
\nonumber
\\
&= A^{\sm(0)} e^{i k_\beta x'^\beta+ih\tilde\varphi} \left(                                                              i k                            \begin{Bmatrix} \cos\theta \cos\phi \\ \cos\theta\sin\phi\\ -\sin\theta \end{Bmatrix}  \right.
                                                                                                                  &            +  \frac{i}{2}  h k  \cos[K(-t+z)]   \begin{Bmatrix} -(1+\cos\theta)\cos\phi \\  (1+\cos\theta)\sin\phi \\  \sin\theta \cos[2\phi] \end{Bmatrix} \quad\ 
                                                                                        \nonumber \\&   &\left.      + \frac{1}{2} h K     \sin[K(-t+z)]    \begin{Bmatrix}   \cos\phi\\ -\sin\phi\\ 0\end{Bmatrix} 
                                                                                   \right).\nonumber
\end{alignat}
Although the fields presented are in the local Lorentz frame of the observer, the parameters $\{\theta,\phi\}$ refer to the underlying zero-order wave, given by $a^{\sm(0)}$. 
Over the time span of one observation, likely much shorter than $1/K$, the observer will quite possibly refrence the observed fields to the short-term observed wavevector,
which is opposite the source direction $\bm{\hat s}'$ found in Section\ \ref{sec:ApparentSourceDirection}.
The fields for the TM mode are related to these by the duality transformation, Equation\ \ref{eq:DualTensorDefs}.
Alternatively, the TM fields can be calculated using the candidate vector potential Equation\ \ref{eq:aCandidateVectorPotentialDef} with Equations\ \ref{eq:qdalphaTMdef} and\ \ref{eq:nualdefTM},
leading to the same result.

\subsection{Poynting 3-vector and direction}\label{sec:Poynting}

The Poynting 3-vector for the linearly-polarized wave above, without averaging, is a set of propagating nearly-plane waves of Poynting flux, with period $\pi/k$.
An average over many oscillations of the electromagnetic field provides the wave direction and amplitude.
Through first order in $h$, this average is:
\begin{align}
\left\langle \bm{S} \right\rangle_{2 \pi N/k} 
&=\left\langle \mathrm{Re}\left[ \bm{E}^{\sm(\mathrm{E})} \right] \times\mathrm{Re}\left[  \bm{B}^{\sm(\mathrm{E})} \right] \right\rangle_{2 \pi N/k} \\
&=
\frac{1}{2} \left| A^{\sm(0)} \right|^2 \left(
	k^2                                        \begin{Bmatrix} \sin\theta\cos\phi                                                     \\         \sin\theta\sin\phi                                                    \\  \cos\theta                                  \end{Bmatrix} 
	+h \frac{k^2}{2}\cos[K(-t+z)] \begin{Bmatrix} -\sin\theta \cos\phi ( 1 + \cos[2\phi](1+\cos\theta))  \\  \PM \sin\theta \sin\phi ( 1 - \cos[2\phi](1+\cos\theta))  \\  - \cos[2\phi] (1+\cos\theta)^2  \end{Bmatrix} \right)
,	\nonumber
\end{align}
where the subscripted angular brackets $\langle ... \rangle_{2 \pi N/k}$ indicate an average over many cycles of the electromagnetic field
and $^*$ indicates the complex conjugate.
Note that the first-order fields proportional to $h \sin[K(-t+z)]$ do not contribute to $\bm{S}$ through first order, because they are at phase quadrature with the 0-order fields, by the factor $i$ in Equation\ \ref{eq:TE_symbolic}.
The duality transformation leads to the same Poynting vector for the TM modes.

The apparent direction of the source is the negative of the normalized Poynting 3-vector:
\begin{align}
\bm{\hat s} &= \frac{1}{|\bm S|} \bm{S}
\end{align}
It is easy to show that this yields the same result as the scalar field, given in Section\ \ref{sec:ApparentSourceDirection}.

The intensity is the magnitude of the Poynting vector:
\begin{align}
\label{eq:TotalIntensityFromPoynting}
|\bm{S}| &= \haf k^2 \left| A^{\sm(0)} \right|^2 \left(1 -  h (1+\cos\theta)\cos[2 \phi] \cos[K(-t+z)] \right)
\end{align}
This expression is particularly interesting because it is not consistent with the expectation from geometric optics, 
that any intensity change produced by a gravitational wave in Minkowski spacetime should be proportional to $h^2$ \cite{zipoy1968reply}.
In this case, solution of the electromagnetic field equations yields a change of intensity proportional to $h$.
However, as discussed in Sections\ \ref{sec:WaveEqScalarField},\ \ref{sec:FieldTensor}, and\ \ref{sec:GAandSolutionForCandidate},
the interaction of gravitational wave and electromagnetic wave results in four scattered waves, 
and is diffractive. 
It is not subject to the rules of geometric optics.

\section{Discussion and Summary}\label{sec:Summary+Discussion}

\subsection{Discussion}\label{sec:Discussion}

The most salient result from the standpoint of potential observations of this work is the change in intensity of the electromagnetic plane wave in gravitational-wave spacetime,
Equation\ \ref{eq:TotalIntensityFromPoynting}. 
This is a diffractive effect and is not subject to the principles of geometric optics.
The process is analogous to diffraction from a transmission grating, or Brillouin scattering of light from sound waves in matter.
The amplitude variations result from the interference of the scattered waves, in particular those represented by the two waves that comprise $h \tilde f q_\alpha$, with the zero-order solution.

A change in the intensity is related to a change in the flux of energy. In a wave field, the energy is not constrained to within ray paths; this holds only in the geometric-optics limit.
The energy flux predicted by Equation\ \ref{eq:TotalIntensityFromPoynting} averages out over many periods of the gravitational wave,
or over many gravitational-wave wavelengths perpendicular to the direction of the electromagnetic wave,
as predicted by the spacetime dependence $\cos[K(-t+z)]$.

Observational tests to verify the change in intensity of electromagnetic waves in gravitational-wave spacetime would be interesting, but quite difficult.
A network of spacecraft, extending over many AU, carrying atomic clocks and linked by phase-locked lasers,
could in principle detect the effect \cite{Eubanks2018DeepSpaceClocks}.
Detection of the predicted polarization variations may be easier than precise measurements of intensity variations.
Such networks are possible, but not yet technically feasible.
Signals from spacecraft do not propagate as perfect plane waves over AU distances, so this work would require extension to more complicated geometry for the electromagnetic waves.

The effect may be more important for environments where gravitational waves are stronger, 
such as the early universe or in the neighborhoods of close binary stars or black holes.
For the early universe, one could compare the present results with those from geometric optics, which have been used to constrain the energy density of gravitational waves in the early universe \cite{Linder1988ApJ...326..517L,Linder1988ApJ...328...77L}.
In principle, interstellar scattering provides access to the close environments of some binary stars, offering accurate phase measurements over scales of an effective instrument spanning many AU \cite{YangPenTestingGravityScintillation2017}.
Again, such studies requier extending the perturbed Minkowski spacetime of the present paper to more complicated geometry for the unperturbed spacetime.

A major goal of the present work is to lay groundwork for further studies. These include extensions of the unperturbed spacetime geometry to more general cases, and extensions of the unpertubed scalar and electromagnetic waves to cases other than plane waves.
The assumption of plane electromagnetic waves is particularly restrictive.
The coincidence of the phase chage $h \tilde \varphi$ with geometric-optics might reasonably be expected, 
because one formulation of the ray approximation states that light travels along the path of stationary phase
\citep{born2013principles}.
The tangent vector to the ray path is then normal to the local gradient of phase at each event.
For plane waves, that gradient is of course the locally-measured direction of the wave.
Forms other than plane waves for the gravitational wave, as investigated by Kopeikin et al.\ \cite{Kopeikin99,1999PhRvD..60l4002K,2002PhRvD..65f4025K,2006CQGra..23.4299K,2014frc1.book.....K}, are also crucial to further work.

Among the particular formal aspects of the present work that may be useful for future work are the use of the zero-order solution to form an inhomogeneous wave equation for the first-order solution,
as exhibited in Equations\ \ref{eq:ScalarWaveEqInhomogForm},\ \ref{eq:LorenzGaugeConditionCovariantPerturbativeDuo1}, and \ref{eq:AbstractSourceTermsForEMWave}.
A consequence of these equations is that the scalar wave equation yields the greatest part of the first-order solution, larger than the remaining parts by a factor of $K/K$.

\subsection{Summary}\label{sec:Summary}

This work examines the effects of propagation of a nearly-plane electromagnetic wave through the spacetime of a plane gravitational wave,
from the standpoint of Maxwell's equations.
I assume that the dimensionless strain amplitude of the gravitational wave is small: $h\ll 1$,
so that solutions are through first order in $h$.
This propagation changes the phase, amplitude, and polarization of the electromagnetic wave,
with periodicity of the gravitational wave. 
From the phase change, I find the change in delay and apparent direction found from calculation of the null geodesics in this spacetime.

As an initial case, I consider propagation of a scalar wave through gravitational-wave spacetime in Section\ \ref{sec:WaveEqScalarField},
and find that the gravitational wave introduces two additional, scattered waves,
in a process similar to diffraction by a transmission grating, or Brillouin scattering by sound waves in matter.
These process are all diffractive, and are not well described by geometric optics.
However, the scattered waves for a scalar field in gravitational-wave spacetime are well described by a varying phase offset
$h \tilde \varphi$ (Equation\ \ref{eq:ScalarVarphidef}).
The change of delay resulting from $h \tilde \varphi$ is that found from calculations using geodesics in earlier work.

I then consider Maxwell's equations and the Lorenz gauge criterion for the electromagnetic 4-vector potential $a^\alpha$, in the spacetime of the gravitational wave, in Section\ \ref{sec:EMinGWspacetime}.
I consider the equations in the absence of charges and currents.
For the solution to zero order in $h$, $a^{\alpha\, \sm(0)}$,
the equations lead to the well-known homogeneous wave equation for each component of $a^{\alpha\, \sm(0)}$, and the Lorenz gauge condition, that hold in Minkowski spacetime.
For the spacetime of a gravitational wave, corrections of order $h$ lead to an inhomogeneous wave equation for each component of the first-order correction $h a^{\alpha\, \sm(1)}$, 
where the differences from the Minkowski-space metric effectively produce an extended effective source of order $h$, involving the zero-order solution.

I consider the case of electromagnetic plane waves in Section\ \ref{sec:NearlyPlaneEMWaves}.
I define polarization basis states (TE and TM) for $a^{\alpha\, \sm(0)}$,
and solve the equations of Section\ \ref{sec:EMinGWspacetime} to find $a^{\alpha\, \sm(1)}$. 
This solution requires at least four scattered waves to represent $h a^{\alpha\, \sm(1)}$.
Two of these scattered waves are identical to the results for the scalar wave, and can be represented by the varying phase $h \tilde \varphi$.
The other two scattered waves mix the components of $a^{\alpha\, \sm(0)}$ in the effective extended source.
They are real rather than imaginary, and thus produce variations of amplitude and polarization.
These variations are smaller than the phase variations by a factor of $K/k$, where $K$ is the frequency of the gravitational wave. In practice $K/k\ll 1$, so these corrections are quite small. 

I calculate observables in Sections\ \ref{sec:DelayDirectionScalar} and\ \ref{sec:DelayDirectionEM}.
Observables are calculated in the local Lorentz frame of the observer.
The observed delay and apparent deflection of a scalar wave are calculated from the wave phase, and are identical to those calculated from the geodesic.
The phase of the electromagnetic wave is the same, and leads to the same observed delay.
The direction is most easily calculated from the electromagnetic field in the observer's local Lorentz frame, and the normalized Poynting 3-vector.
The deflection is the same as that for a scalar field.
However, the magnitude of the Poynting vector, the intensity of the electromagnetic wave, changes with time and position because of interference between the scattered waves and the zero-order plane wave.
This produces periodic variations in intensity.
In Section\ \ref{sec:Discussion}, I describe possible observations of the predicted amplitude change, and the other observables,
and point out that such tests will require calculations with more realistic geometry for the unperturbed spacetime and electromagnetic wave, than the simple case of Minkowski spacetime with electromagnetic plane waves considered here.

\section*{Acknowledgements}

I thank Jim Hartle, Mark Srednicki, Sergei Kopeikin, and Marshall Eubanks for useful discussions. I thank Xinzhong Er and Michael Rupen for re-stimulating my interest in this problem,
and Ted Pyne and Irwin Shapiro for their encouragement.

\bibliographystyle{jphysicsB}
\bibliography{CRGwinn_2} %

\end{document}